\documentclass[12pt,draftclsnofoot,onecolumn]{IEEEtran}

\usepackage{amsmath,graphics,amssymb,epsfig,subfigure,color,cite}
\usepackage{array}
\usepackage{multirow}
\usepackage{enumerate}
\usepackage{algorithmic}
\usepackage{algorithm}
\usepackage{bm}
\usepackage{float}
\usepackage{makecell}

\usepackage{lipsum}
\usepackage{capt-of}

\newtheorem{thm}{Theorem}

\newtheorem{prop}{Proposition}

\floatname{algorithm}{Procedure}

\makeatletter
\newcommand{\vast}{\bBigg@{4.5}}
\newcommand{\Vast}{\bBigg@{7.5}}
\makeatother

\begin{document}
    \title{Communication-Efficient Federated Learning via Quantized Compressed Sensing}
	%\title{FedQCS: Federated Learning via \\  Quantized Compressed Sensing} <- For TSP,
	%\title{Quantized Compressed Sensing for Communication-Efficient Federated Learning}

	\author{Yongjeong Oh, Namyoon Lee, Yo-Seb Jeon, and H. Vincent Poor
	    \thanks{This paper will be presented in part at the 2021 IEEE Global Communications Conference Workshops \cite{Conference}.}
		\thanks{Y. Oh, N. Lee, and Y.-S. Jeon are with the Department of Electrical Engineering, POSTECH, Pohang, Gyeongbuk 37673, South Korea (e-mails: \{yongjeongoh,nylee,yoseb.jeon\}@postech.ac.kr).}
		%\thanks{M. M. Amiri is with the Department of Electrical Engineering, Princeton University, Princeton, NJ 08544 (e-mails: mamiri@princeton.edu).}
		\thanks{H. V. Poor is with the Department of Electrical and Computer Engineering, Princeton University, Princeton, NJ 08544 (e-mail: poor@princeton.edu).}
		%\thanks{This work was supported in part by the U.S. National Science Foundation under Grants  CCF-0939370 and CCF-1908308.} %This work was supported in part by Samsung Research Funding $\&$ Incubation Center of Samsung Electronics under Project Number SRFC-IT1702-00, and in part by the National Science Foundation under Grant No. NSF-CCF-1527079. This work was presented in part at the 2018 IEEE 87th Vehicular Techonology Conference (VTC2018-Spring).}
	}
	\vspace{-2mm}	
	
	\maketitle
	\vspace{-12mm}

	\begin{abstract} % up to 200 words
	    In this paper, we present a communication-efficient federated learning framework inspired by quantized compressed sensing. The presented framework consists of gradient compression for wireless devices and gradient reconstruction for a parameter server (PS). Our strategy for gradient compression is to sequentially perform block sparsification, dimensional reduction, and quantization. Thanks to gradient sparsification and quantization, our strategy can achieve a higher compression ratio than one-bit gradient compression. For accurate aggregation of the local gradients from the compressed signals at the PS, we put forth an approximate minimum mean square error (MMSE) approach for gradient reconstruction using the expectation-maximization generalized-approximate-message-passing (EM-GAMP) algorithm. Assuming Bernoulli Gaussian-mixture prior, this algorithm iteratively updates the posterior mean and variance of local gradients from the compressed signals. We also present a low-complexity approach for the gradient reconstruction. In this approach, we use the Bussgang theorem to aggregate local gradients from the compressed signals, then compute an approximate MMSE estimate of the aggregated gradient using the EM-GAMP algorithm. We also provide a convergence rate analysis of the presented framework. Using the MNIST dataset, we demonstrate that the presented framework achieves almost identical performance with the case that performs no compression, while significantly reducing communication overhead for federated learning.
	\end{abstract}


	\section{Introduction}\label{Sec:Intro}
	Federated learning is a decentralized artificial intelligence (AI) technique for training a {\em global} model on a parameter server (PS) through collaboration with wireless devices, each with its own local training dataset \cite{Konecny:15,McMahan:17,Niknam:19,Zhu:20,Gunduz:20}. 
	The most widely adopted approach in federated learning is to update the global model by iterating the following two steps:
	%  In federated learning, the devices do not directly transmit their training data samples to the PS; instead, they collaborate with the PS in 
	(i) each wireless device updates a \textit{local} model based on its local training dataset, then transmits the information of the local model update; (ii) the PS updates the global model by aggregating the local model updates transmitted by the devices, then broadcasts the updated model to the devices. 
	Federated learning based on the above approach allows the PS to train the global model without direct access to the devices' data and therefore can help preserve the privacy of the data generated at the devices.
	%By continuing the collaboration between the PS and the devices, the PS is able to train the global model without no direct access to the devices' data.
	Thanks to this advantage, federated learning has received a great deal of attention as a means of enabling privacy-sensitive AI applications \cite{Niknam:19,Zhu:20,Gunduz:20,Samarakoon:19,Chen:21,Jeon:21}.

	%In federated learning, each wireless device updates a \textit{local} model based on its local training data set and then sends the local model update (e.g., a gradient vector of the local model) to the PS. 
	%The PS then updates a \textit{global} model by aggregating the local model updates and then broadcasts the updated global model to the devices. 
	%By repeating this collaboration between the PS and the devices, the PS is able to train the global model without directly accessing the devices' data, thereby enhancing the privacy of data generated at the devices.   
	%

	%A major problem in federated learning is that transmitting the local model updates from the wireless devices to the PS requires significant communication overhead that may not be affordable in practical communication systems. 
	A major bottleneck in federated learning is significant communication overhead required when transmitting the local model updates from the wireless devices to the PS.
	This problem becomes more severe as the global model on the PS becomes more sophisticated, because the amount of the communication overhead increases with the number of global model parameters. 
	To address this problem, gradient compression for federated learning is necessary, and several compression techniques have been intensively proposed in the literature  \cite{Konecny:16,Alistarh:17,SignSGD,Lee:21,UVeqFed,Du:20,Aji:17,Wangni:18,Lin:18,Amiri:TSP,Amiri:TWC,UVeqFed2,Hwang:21}.   
	%gradient compression techniques for federated learning have been intensively studied in the literature \cite{Konecny:16,Alistarh:17,SignSGD,Lee:21,UVeqFed,Du:20,Aji:17,Wangni:18,Lin:18,Amiri:TSP,Amiri:TWC}.  
	The common idea of these techniques is to apply lossy compression to the local gradients of the model parameters computed at each device. 
	%the gradients of the model parameters computed at each device based on its local data set. 
	Two representative approaches for gradient compression are \textit{gradient quantization} and \textit{gradient sparsification}.
	%{\em Gradient quantization:}
	In the gradient quantization approach, local gradients are quantized and then transmitted using digital transmission \cite{Konecny:16,Alistarh:17,SignSGD,Lee:21,UVeqFed,Du:20}. 
	A well-known technique based on this approach is one-bit quantization, in which the device only transmits the sign of each local gradient \cite{SignSGD}. 
	Vector quantization for gradient compression is also studied in \cite{UVeqFed,Du:20}.
	In these studies, the local gradients are partitioned into multiple groups; each group of the local gradients is quantized using vector quantizers such as lattice quantizers \cite{UVeqFed} and Grassmannian quantizers \cite{Du:20}. 
	%A common limitation of these techniques is that the communication overhead of federated learning cannot be made less than the number of the model parameters.
	%{\em Gradient sparsification:}
	In the gradient sparsification approach, local gradients are sparsified by dropping less significant entries in their magnitudes.
	Analog transmission with gradient sparsification is studied in \cite{Amiri:TSP,Amiri:TWC} in which a local gradient vector after the sparsification is compressed by random projection onto a lower dimensional space as in compressed sensing (CS). 
	Digital transmission with gradient sparsification is studied in \cite{Aji:17,Wangni:18,Lin:18} which an encoding function is designed to exploit the sparsity of the local gradients.
	%Despite these efforts, reduction in the communication overhead achieved by the gradient sparsification approach is limited in general compared to the gradient quantization approach. 

	Recently, gradient compression based on quantized compressed sensing (QCS) has been studied in \cite{DitherQCS,QCS1,QCS2} to take the advantages of both the gradient quantization and sparsification. 
	A representative strategy towards this direction is gradient compression based on QCS with dithered uniform quantization \cite{DitherQCS}.  
	A key advantage of this strategy is that quantization error can be transformed into an independent noise, which allows for the PS to reconstruct the local gradients using a simple linear estimator. 
	This advantage, however, is attained at the cost of additional communication overhead because each device needs to additionally transmit its dither signal to the PS, which scales linearly with the number of global model parameters. 
	Gradient compression based on QCS without dithered quantization is studied in \cite{QCS1,QCS2}, in which binary iterative hard thresholding algorithm is adopted to reconstruct gradients at the PS.
	A common limitation of the strategies in \cite{QCS1,QCS2} is that they only operate with one-bit quantization and therefore suffer from high quantization error. 
	Despite these efforts, none of the existing studies has developed gradient compression based on QCS with a multi-bit non-uniform scalar quantizer which has more flexibility to control the compression ratio and the quantization noise level than the existing work. %which has a better potential to reduce quantization error than one-bit quantization or uniform quantization.
	More importantly, minimizing reconstruction error at the PS remains unsolved in federated learning with QCS-based gradient compression, which is essential to reduce the performance gap between centralized learning and federated learning.  
% 	The use of QCS has a potential to achieve significant reduction in the communication overhead of federated learning, but reconstruction of the gradients at the PS becomes more challenging due to the presence of both quantization and compression errors.  
% 	To overcome this challenge, gradient compression based on dithered quantization with compressed sensing is studied in \cite{DitherQCS}.
% 	A key advantage of the dithered quantization is that it simplifies gradient reconstruction at the PS as quantization error can be transformed into an independent noise when adding random dither signals before the scalar quantization \cite{DitherTheory}. 
% 	This advantage, however, is attained at the cost of additional communication overhead because each device needs to transmit its dither signal to the PS.
% 	Gradient compression based on one-bit compressed sensing is also studied in \cite{QCS1,QCS2} in which binary iterative hard thresholding (BIHT) algorithm in \cite{BIHT} is adopted for the gradient reconstruction at the PS. 
% 	The common limitation of these studies is that they do not support QCS with multi-bit quantization which has a better potential to improve the performance of federated learning.
% 	To the authors' best knowledge, none of the existing studies have developed the QCS-based gradient compression with general scalar quantization.
% 	Moreover, minimizing the gradient reconstruction error for the QCS-based gradient compression is still an open problem, which is essential to reduce the performance gap between centralized learning and federated learning.   

	In this paper, we present a communication-efficient federated learning framework, referred to as {\em FedQCS}. 
	This framework consists of a gradient compression strategy for wireless devices and gradient reconstruction strategies for a PS. 
	Our gradient compression strategy, inspired by QCS, effectively reduces communication overhead of transmitting a local gradient vector from each device to the PS, by taking the advantages of both the gradient quantization and sparsification. 
	Meanwhile, our gradient reconstruction strategies enable accurate aggregation of the local gradients from the compressed signals at the PS. 
	We also provide a convergence rate analysis of FedQCS. 
	Using the MNIST dataset, we demonstrate that FedQCS with one bit overhead per gradient entry performs very close to the case with no compression, while outperforming the existing QCS-based federated learning frameworks.
% 	We first develop a gradient compression strategy for the wireless devices, which not only achieves significant reduction in the communication overhead, but also enables the design of the optimal scalar quantizer. 
% 	We then develop two gradient reconstruction strategies for the PS in FedQCS, which effectively reduce gradient reconstruction error at the PS by leveraging novel variants of the expectation-maximization generalized-approximate-message-passing (EM-GAMP) algorithm in \cite{EMGAMP}.
% 	We also analyze the reconstruction error as well as the convergence rate of FedQCS. 
% 	Using simulations, we demonstrate that FedQCS enables almost lossless gradient reconstruction at the PS while achieving significant reduction in the communication overhead. 
	The major contributions of this paper are summarized as follows:
	\begin{itemize}
		\item We propose a gradient compression strategy to reduce communication overhead of transmitting a local gradient vector from each device to the PS. 
		The key idea of our strategy is to sequentially perform (i) block sparsification, (ii) dimension reduction, and (iii) quantization. 
		%The key idea of the proposed strategy is to sparsify the local gradient vector in a block-wise fashion, then compress-and-quantize each block of the sparsified gradient vector.
		In the block sparsification process, we divide the local gradient vector at each device into $B$ sub-vectors, then sparsifies each sub-vector by dropping the least significant entries in their magnitudes. 
		Then, in the dimension reduction process, we reduce the dimension of each sparsified sub-vector by applying random projection onto a lower dimensional space using a sensing matrix. 
		Finally, in the quantization process, each entry of the low-dimensional sub-vector is quantized by using the optimal Llyod-Max scalar quantizer. 
		It is demonstrated that our compression strategy requires $\frac{Q}{R}$ bits for conveying the information of each local gradient entry to the PS, where $Q$ is the quantization bits of the scalar quantizer and $R$ is a dimension reduction ratio. 
		Therefore, our strategy not only provides a flexible communication overhead for federated learning, but also achieves a higher compression ratio than the state-of-the-art one-bit gradient compression (e.g., \cite{SignSGD}) when $Q<R$.
		%By taking the advantages of both gradient sparsification and quantization, our compression strategy achieves a considerable reduction in communication overhead for federated learning. 
		%For example, flexible communication overhead, given by $\frac{Q}{R}$ bits per gradient entry, where $Q$ is the quantization bits of the scalar quantizer and $R$ is a dimension reduction ratio. 
		%provides a flexible communication overhead, given by $\frac{Q}{R}$ bits per gradient entry, where $Q$ is the quantization bits of the scalar quantizer and $R$ is a dimension reduction ratio. %
		%The most prominent feature of our compression strategy is that provides
		%For example, by choosing $Q<R$, our strategy achieves a higher compression ratio than the state-of-the-art one-bit gradient compression \cite{SignSGD}.
		%Thanks to this feature, our strategy Thanks to this feature, our strategy with $Q<R$ achieves a higher compression ratio than the state-of-the-art one-bit gradient compression \cite{SignSGD}.
		
		%Our strategy also enables the design of the optimal scalar quantizer because every input of the scalar quantizer follows a Gaussian distribution with zero mean and unit variance.
		
		\item 
		We develop two gradient reconstruction strategies for the PS, referred to as {\em estimate-and-aggregate} and {\em aggregate-and-estimate}, which enable accurate aggregation of local gradients from compressed signals. 
		The key idea of the estimate-and-aggregate strategy is to first estimate each local gradient sub-vector from its compressed signal, then aggregate the estimated sub-vectors to reconstruct a global gradient vector. 
		In this strategy, the problem of estimating each local gradient sub-vector is formulated as a \textit{quantized} CS recovery problem. 
        We solve this problem by employing a quantized variant of the expectation-maximization generalized-approximate-message-passing (EM-GAMP) algorithm in \cite{EMGAMP,QEMGAMP} with Bernoulli Gaussian-mixture prior, which iteratively computes an approximate minimum mean square error (MMSE) estimate of the local gradient sub-vector from the compressed signal.
		Although the estimate-and-aggregate strategy approximately minimizes the MSE of the local gradient estimates, the computational complexity of this strategy increases linearly with the number of the devices.
		To mitigate this complexity requirement, in the aggregate-and-estimate strategy, we first aggregate the local gradient sub-vectors and then estimate the aggregated sub-vector. 
		The underlying challenge in this strategy is that aggregation of the local gradient sub-vectors from the compressed signals is not straightforward due to nonlinearity of the quantization.
		To overcome this challenge, we use the Bussgang theorem in \cite{Bussgang} to transform a nonlinear compressed signal into a linear compressed signal with additive distortion. 
		Thanks to this theorem, the problem of estimating the aggregated sub-vector is formulated as an \textit{unquantized but noisy} CS recovery problem. 
		We solve this problem by employing the original EM-GAMP algorithm in \cite{EMGAMP}, which iteratively computes an approximate MMSE estimate of the aggregated sub-vector from its noisy linear observation. 
		A key advantage of the aggregate-and-estimate strategy is that it can adjust the performance-complexity trade-off of the gradient reconstruction process by changing how many sub-vectors are aggregated before the estimation.

		\item 
		We provide a convergence rate analysis of FedQCS using the aggregate-and-estimate strategy.
		%We analyze the gradient reconstruction error as well as the convergence rate of FedQCS when employing the joint recovery strategy. 
		%For the reconstruction error analysis, we assume that every local gradient vector follows a Bernoulli Gaussian-mixture model which is known at the PS in prior. 
		To this end, we first characterize an MSE upper bound in reconstructing a global gradient vector based on the Bernoulli Gaussian-mixture modeling of the local gradient vector.
		%characterize an upper bound for the reconstruction error achieved by FedQCS with the aggregate-and-estimate strategy in Sec.~\ref{Sec:BussGAMP}.
        %We then characterize the convergence rate of FedQCS under the consideration of the reconstruction error bound. 
		Our analysis demonstrates that the reconstruction error reduces as a dimension reduction ratio, $R$, decreases and also as the number of the quantization bits, $Q$, increases.
		We then use the reconstruction error bound to characterize the convergence rate of FedQCS operating with a stochastic gradient descent (SGD) algorithm.
		From the analysis, we show that FedQCS is guaranteed to converge to a stationary point of a smooth loss function at the rate of $\mathcal{O}\big(\frac{1}{\sqrt{T}}\big)$, where $T$ is the number of total iterations of the SGD algorithm.   
		%Using the derived MSE bound. show that FedQCS is guaranteed to converge to a stationary point of a smooth loss function at the rate of $\mathcal{O}\big(\frac{1}{\sqrt{T}}\big)$, where $T$ is the number of total iterations of the SGD algorithm.   
		% then characterize the convergence rate of FedQCS operating with a stochastic gradient descent (SGD) algorithm.		Our result 

		\item Using simulations, we demonstrate the superiority of FedQCS over the existing QCS-based federated learning frameworks for an image classification task using the MNIST dataset \cite{MNIST}.
		%In these simulations, we compare the classification accuracy and the normalized MSE of the proposed algorithm with those of conventional compressed sensing algorithms.
		Our simulation results demonstrate that FedQCS with one bit overhead per gradient entry suffices to attain the identical classification accuracy as perfect reconstruction with no compression. 
		It is also shown that FedQCS outperforms the existing QCS-based frameworks in terms of both the classification accuracy and the normalized MSE of the gradient reconstruction.
		%It is also shown that the performance-complexity trade-off of the proposed algorithm can be adjusted by employing the group-wise estimation strategy.
		We also investigate the effect of the communication overhead, the dimension reduction ratio, the number of the quantization bits, and the sparsification level on the performance of FedQCS.
		From simulation results, we demonstrate that FedQCS effectively reduces communication overhead of federated learning while enabling accurate reconstruction of the global gradient vector at the PS.
		%, by taking the advantages of both the gradient quantization and sparsification.
		%Based on these simulation results, we demonstrate that FedQCS enables almost lossless gradient reconstruction at the PS while  reducing the communication overhead by taking the advantages of both the gradient quantization and sparsification.

	\end{itemize}

	\subsubsection*{Notation}
	Upper-case and lower-case boldface letters denote matrices and column vectors, respectively.
	$\mathbb{E}[\cdot]$ is the statistical expectation,
	%$\mathbb{P}(\cdot)$ is the probability,
	and $(\cdot)^{\sf T}$ is the transpose.
	%and $(\cdot)^{\sf H}$ is the conjugate transpose,
% 	$\lceil \cdot \rceil$ is the ceiling function,
% 	$\lfloor \cdot \rfloor$ is the floor function,
%	and $|\cdot|$ is the absolute value.
% 	${\sf Re}\{\cdot\}$ and ${\sf Im}\{\cdot\}$ denote real and imaginary components, respectively.
	$|\mathcal{A}|$ is the cardinality of set $\mathcal{A}$.
	$({\bf a})_i$ represents the $i$-th entry of vector ${\bf a}$.
	$\|{\bf a}\|\!=\!\sqrt{{\bf a}^{\sf T}{\bf a}}$ is the Euclidean norm of a real vector ${\bf a}$.
	%and $\|{\bf A}\|_{\rm F}\!=\!\sqrt{{\sf Tr}({\bf A}{\bf A}^{\sf H})}$ is the Frobenius norm of a matrix ${\bf A}$.
	%$\mathbb{I}\{\mathcal{A}\}$ is an indicator function which equals one if an event $\mathcal{A}$ is true and zero otherwise.
	$\mathcal{N}({\bm \mu},{\bf R})$ represents the distribution of a Gaussian random vector with mean vector ${\bm \mu}$ and covariance matrix ${\bf R}$.
	${\bf 0}_n$ is an $n$-dimensional vector with zero entries.
	${\bf I}_N$ is an $N$ by $N$ identity matrix.
% 	$\mathbb{R}$ is the set of real numbers.

% 	\begin{figure*}[t]
% 		\centering 
% 		{\epsfig{file=Figures/Fig_System.eps, width=16cm}}
% 		\caption{An illustration of uplink communication of federated learning over a wireless MIMO multiple access channel.} \vspace{-7mm}
% 		\label{fig:System}
% 	\end{figure*}

	\section{System Model}\label{Sec:Model}
	%In this section, we present the system model considered in this work. We then introduce a gradient estimation problem in the considered system.	
% 	%%%%%%%%%%%%%%%%%%%%%%%%%%%%%%%%%%%%%%%%%%%%%%%%%%%%%%%%%%%%%%%%%%%%%%%%%%%%%%%%%%%%%%%%
% 	%%%%%%%%%%%%%%%%%%%%%%%%%%%%%%%%%%%%%%%%%%%%%%%%%%%%%%%%%%%%%%%%%%%%%%%%%%%%%%%%%%%%%%%%
% 	\subsection{Federated Learning over a Wireless MIMO MAC}	
	We consider a federated learning scenario in which a global model on a parameter server (PS) is trained by collaborating with $K$ wireless devices.
	A key assumption in federated learning is that data samples for training the global model are distributed over the wireless devices, while the PS has no direct access to these samples. 
	We denote a set of training data samples available at device $k\in \mathcal{K}=\{1,\ldots,K\}$ by $\mathcal{D}_k$, which is hereafter referred to as a {\em local} training dataset.
	We also denote a parameter vector that represents the global model on the PS by ${\bf w}\in \mathbb{R}^{\bar{N}}$, where $\bar{N}$ is the number of the parameters.
	%each with its own local training data set
	%The global model can be represented by a parameter vector , where $\bar{N}$ is the number of the parameters.
	For example, if the global model takes a form of a deep neural network (DNN), the entries of the parameter vector are the weights and the biases of the DNN. 
	%Let $\mathcal{D}_k$ be a local training data set available at device $k$, which consists of $|\mathcal{D}_k|$ training data samples for $k\in \mathcal{K}=\{1,\ldots,K\}$.
	Then a \textit{local} loss function at device $k$ for the parameter vector ${\bf w}$ is defined as 
	\begin{align}
	    F_k({\bf w}) = \frac{1}{|\mathcal{D}_k|} \sum_{{\bf u}\in\mathcal{D}_k} f({\bf w};{\bf u}),
	\end{align}
	where $f({\bf w};{\bf u})$ is a loss function computed for the parameter vector ${\bf w}$ with respect to a training data sample ${\bf u}\in\mathcal{D}_k$.
	Similarly, a \textit{global} loss function for the parameter vector ${\bf w}$ is defined as
	\begin{align}\label{eq:global_loss}
	    F({\bf w}) = \frac{1}{|\mathcal{D}|} 
	    \sum_{{\bf u}\in \mathcal{D}} f({\bf w};{\bf u})
	    = \frac{1}{\sum_{j=1}^K|\mathcal{D}_j|} \sum_{k=1}^K  |\mathcal{D}_k|  F_k({\bf w}),
	\end{align}
	where $\mathcal{D} = \cup_k \mathcal{D}_k$.
	The ultimate goal of federated learning is to find the best parameter vector that minimizes the global loss function in \eqref{eq:global_loss}.
	A practical solution to achieve this goal is to train the parameter vector based on a gradient-based optimizer such as a stochastic gradient descent algorithm and the ADAM optimizer in \cite{ADAM}.
	Let ${\bf w}_t\in{\mathbb R}^{\bar{N}}$ be the parameter vector at iteration $t \in \{1,\ldots, T\}$ of the optimizer, where $T$ is the total number of iterations.
% 	A common requirement of the gradient-based optimizer is the information of a global gradient vector defined as 
% 	\begin{align}\label{eq:global_grad}
% 	    \nabla f({\bf w};\mathcal{D}_{\rm tot}) = \frac{1}{|\mathcal{D}_{\rm tot}|} \sum_{{\bf u}\in\mathcal{D}_{\rm tot}} \nabla f({\bf w};{\bf u}),
% 	\end{align}	
% 	where $\nabla$ is a gradient operator. 
    Then minimizing the loss function in \eqref{eq:global_loss} using the gradient-based optimizer requires the knowledge of a {\em true} gradient vector at the PS, defined as
    \begin{align}\label{eq:true_grad}
        \nabla F ({\bf w}_t) = \frac{1}{|\mathcal{D}|}  \sum_{{\bf u}\in \mathcal{D}} \nabla f({\bf w}_t;{\bf u}),~~\forall t\in\{1,\ldots,T\}.
    \end{align}

	In federated learning, training data samples are available only at the wireless devices; thereby, the true gradient vector in \eqref{eq:true_grad} cannot be computed at the PS directly. 
	As an alternative solution, the PS acquires the knowledge of the gradient vector by collaborating with the wireless devices as described below.

	{\bf Operations at the wireless devices:} 
	Suppose that all the wireless devices have the information of a globally consistent parameter vector ${\bf w}_t$.
	Each wireless device computes a {\em local} gradient vector based on its own local training dataset.
    A local gradient vector computed at device $k$ for the parameter vector ${\bf w}_t$ is given by 
    \begin{align}\label{eq:local_grad0}
	    %\nabla f_k({\bf w}_t;\mathcal{D}_k^{(t)} )
	    \nabla F_{k}^{(t)}\big({\bf w}_t\big) = \frac{1}{|\mathcal{D}_k^{(t)}|} \sum_{{\bf u}\in\mathcal{D}_k^{(t)}} \nabla f({\bf w}_t;{\bf u}),
	\end{align}
	where $\nabla$ is a gradient operator, and $\mathcal{D}_k^{(t)} \subset \mathcal{D}_k$ is a mini-batch randomly drawn from $\mathcal{D}_k$ at iteration $t$. 
	%In \eqref{eq:local_grad}, we assume that the size of the mini-batch is known at the PS and does not change over iterations, i.e., $D_k = |\mathcal{D}_k^{(t)}|$, $\forall t \in \{1,\ldots,T\}$.
	Then all the devices send the information of their local gradient vectors to the PS. 
    Since direct transmission of the local gradient vector in \eqref{eq:local_grad0} imposes large communication overhead when $\bar{N} \gg 1$, we assume that each device applies \textit{lossy} compression to its local gradient vector before the transmission.
    Our strategy for compressing the local gradient vectors will be described in Sec.~\ref{Sec:Comp}.

	{\bf Operations at the parameter server:}
    Based on compressed local gradient vectors sent by the wireless devices, the PS attempts to reconstruct a {\em global} gradient vector defined as 
    \begin{align}\label{eq:global_grad}
	    {\bf g}_{\mathcal{K}}^{(t)} = \sum_{k=1}^K \rho_k^{(t)} \nabla F_{k}^{(t)}\big({\bf w}_t\big) ,
	\end{align}
	where $\rho_k^{(t)} \triangleq \frac{|\mathcal{D}_k^{(t)}|}{\sum_{j=1}^K |\mathcal{D}_j^{(t)}|}$.
	Since the lossy compression is applied at the wireless devices, perfect reconstruction of the global gradient vector is not feasible at the PS.  
	As a result, a global gradient vector reconstructed at the PS, namely $\hat{\bf g}_{\mathcal{K}}^{(t)} \in \mathbb{R}^{\bar{N}}$, contains reconstruction error. 
	Our strategy to minimize this error will be elucidated in Sec.~\ref{Sec:Recovery}.
	After the gradient reconstruction, the PS updates the parameter vector according to the optimizer based on the information of $\hat{\bf g}_{\mathcal{K}}^{(t)}$.
	%Our strategy for reconstructing the global gradient vector will be described in Sec.~\ref{Sec:Recovery}.
	%Since the lossy compression is applied at the wireless devices, perfect reconstruction of the global gradient vector is not feasible at the PS.  
	%As a result, the PS attains a reconstructed global gradient vector, namely $\hat{\bf g}_{\mathcal{K}}^{(t)} \in \mathbb{R}^{\bar{N}}$, which contains reconstruction error. 
	%Then the PS updates the parameter vector according to the optimizer based on the information of $\hat{\bf g}_{\mathcal{K}}^{(t)}$.
	For example, if a gradient descent algorithm is employed at the PS, the corresponding update rule is given by  
	\begin{align}\label{eq:global_update}
	    {\bf w}_{t+1} \leftarrow {\bf w}_t - \eta_t \hat{\bf g}_{\mathcal{K}}^{(t)},
	\end{align}
	where $\eta_t >0$ is a learning rate at iteration $t$. 
	Finally, the PS broadcasts the updated parameter vector to the wireless devices.
	
	\begin{figure*}
    	\centering
    	{\epsfig{file=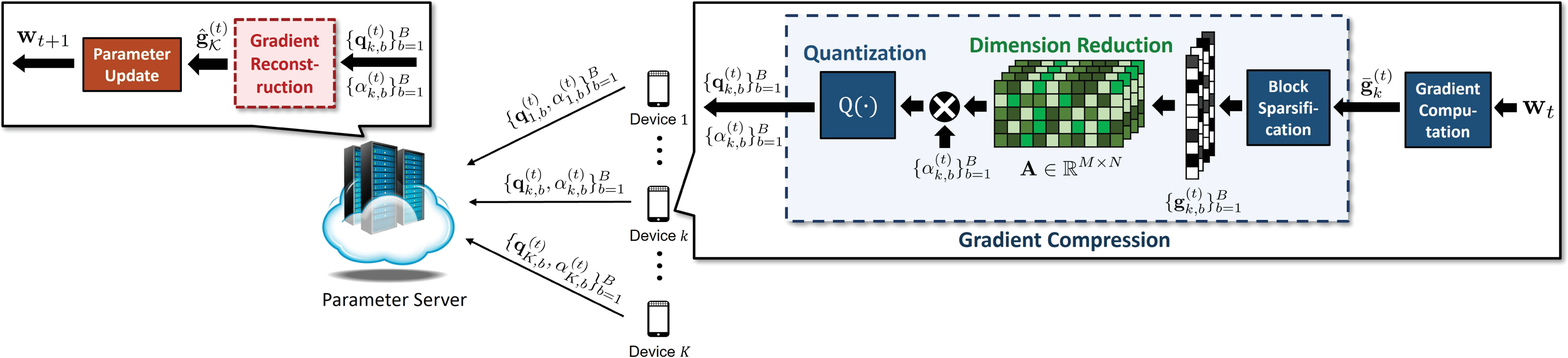, width=16.2cm}}\vspace{-3mm}
    	\caption{An illustration of the proposed federated learning framework.}  \vspace{-3mm}
    	\label{fig:System}
    \end{figure*}

	\vspace{3mm}
	The federated learning scenario described above is illustrated in Fig.~\ref{fig:System}.
	Under this scenario, the major contribution of this work is to present a novel federated learning framework by developing (i) a local gradient compression strategy for the wireless devices and (ii) global gradient reconstruction strategies for the PS.
	We refer to this framework as federated learning via quantized compressed sensing (FedQCS).  
	%borrows the idea from quantized compressed sensing (QCS) for designing both the gradient compression and reconstruction strategies. 
% 	The major focus of this work is to investigate uplink transmission and reception strategies for federated learning; thereby, we assume that the downlink communication is error free. 
%     Under this assumption, all the devices have a globally consistent parameter vector ${\bf w}_t$ for all $t\in\{1,\ldots,T\}$.   

	\section{Gradient Compression Strategy of FedQCS}\label{Sec:Comp}
	Communication overhead reduction in federated learning is essential when optimizing the loss function with a very large model size such as DNNs. In this section, we propose a lossy gradient compression strategy which effectively reduces the communication overhead required when transmitting a local gradient vector from each device to the PS. 
	%A major challenge in federated learning is that direct transmission of local gradient vectors from the wireless devices to the PS requires significant communication overhead when the number of global model parameters is extremely large (i.e., $\bar{N}\gg 1$).
	%To overcome this challenge, in this section, we propose a local gradient compression strategy, inspired by QCS, which effectively reduces the communication overhead required when transmitting a local gradient vector from each device to the PS. 

	\subsection{Proposed Compression Strategy} \label{Sec:Comp_Block}
	The key idea of the proposed strategy, inspired by QCS, is to sequentially perform (i) block sparsification, (ii) dimension reduction, and (iii) quantization. We refer to this strategy as \textit{BQCS} compression as it performs block sparsification before applying the compression based on QCS. 
    The overall procedure of the BQCS compression is summarized in Steps 4--8 in {\bf Procedure~\ref{alg:FedQCS}}, while details of each process are elaborated below. 
    	%The idea of our strategy for local gradient compression is to sequentially perform 1) sparsification, 2) dimension reduction, and 3) quantization. We refer to this strategy as BSDQ compression. 
        	%The key idea of the proposed strategy is to compress-and-quantize a local gradient vector at each device after sparsifying this vector in a block-wise fashion.
        	%Based on this idea, two processes, referred to as {\em block sparsification} and {\em compress-and-quantize}, are successively performed by each wireless device.
	        %Details of the BSDQ compression are elaborated below. 
	
	{\bf Block Sparsification:} 
	In the block sparsification process, each device divides its gradient vector into $B$ sub-vectors, each of which has a dimension of $N = \frac{\bar{N}}{B}$, then sparsifies each sub-vector by dropping the least significant entries in terms of their magnitudes. 
	%To facilitate parallel computing of gradient compression as well as reconstruction processes, each device first divides its \textit{target} local gradient vector into $B$ sub-vectors.
	Let $\bar{\bf g}_{k}^{(t)} \in \mathbb{R}^{N}$ be a \textit{target} local gradient vector computed by device $k$ at iteration $t$. 
    Also, let $\mathcal{N}_1, \ldots, \mathcal{N}_{B}$ be mutually exclusive subsets of $\mathcal{\bar N}=\{1,\ldots,\bar{N}\}$ such that $\bigcup_{b=1}^B \mathcal{N}_b = \mathcal{\bar N}$.
    Then the $b$-th local gradient sub-vector at device $k$ is defined as $\bar{\bf g}_{k,b}^{(t)}
        = [\bar{g}_{k,\mathcal{N}_b(1)}^{(t)}, \cdots, \bar{g}_{k,\mathcal{N}_b(N)}^{(t)} ]^{\sf T}$,
        where $\bar{g}_{k,i}^{(t)}$ is the $i$-th entry of $\bar{\bf g}_{k}^{(t)}$.
    Then an $S$-sparse vector ${\bf g}_{k,b}^{(t)} \in\mathbb{R}^N$ is generated from $\bar{\bf g}_{k,b}^{(t)}$ by dropping all but the top-$S$ entries with the largest magnitudes.
    For ease of exposition, we denote the block sparsification process applied to the local gradient vector $\bar{\bf g}_{k}^{(t)}$ as ${\sf BlockSparse}\big(\bar{\bf g}_{k}^{(t)}\big)$ whose output is given by $\{{\bf g}_{k,b}^{(t)}\}_{b=1}^B$.  
    %A major benefit of the above process is that by choosing $B>1$, each device can facilitate parallel computing of the gradient compression process, which reduces the computational complexity of a gradient reconstruction process performed at the PS. 
	One drawback of the block sparsification process is that gradient information is lost when dropping the least significant gradient entries. 
	Fortunately, this information loss can be partially compensated by accumulating the dropped gradient entries and then by adding these entries in the next iteration \cite{Amiri:TSP,Amiri:TWC}. 
	Motivated by this fact, the block sparsification process is performed in conjunction with gradient accumulation as follows: 
	%Let ${\bf \Delta}_{k,b}^{(t)} \in \mathbb{R}^{N}$ be the accumulated gradient vector stored at device $k$ until iteration $t$, which is initialize as ${\bf \Delta}_{k,b}^{(0)} = {\bf 0}_N$. 
    %Then the accumulated vector at iteration $t+1$ is updated as 
	%\begin{align}
    %    {\bf \Delta}_{k,b}^{(t+1)} = \bar{\bf g}_{k}^{(t)} - {\sf Concatenate}\big(\{{\bf g}_{k,b}^{(t)}\}_{b=1}^B\big),
    %\end{align}	    
    %where ${\sf Concatenate}\big(\{{\bf g}_{k,b}^{(t)}\}_{b=1}^B\big)$ is a function that concatenates $B$ sub-vectors $\{{\bf g}_{k,b}^{(t)}\}_{b=1}^B$ into the form of the original vector ${\bf g}_k^{(t)}$. 
    %The accumulated vectors $\{{\bf \Delta}_{k,b}^{(t+1)}\}_{b=1}^{B}$ are stored at device $k$ and then added to the next local gradient vector at iteration $t+1$: 
 	%\begin{align}\label{eq:local_grad}
    %     \bar{\bf g}_{k}^{(t+1)} = \nabla F_{k}^{(t+1)}\big({\bf w}_{t+1}\big)  + {\bf \Delta}_{k}^{(t+1)}.
    %\end{align}	  
    Let ${\bf \Delta}_{k}^{(t)} \in \mathbb{R}^{\bar{N}}$ be a residual gradient vector of device $k$ at iteration $t+1$, defined as
	\begin{align}
        {\bf \Delta}_{k}^{(t+1)} = \bar{\bf g}_{k}^{(t)} - {\sf Concatenate}\big(\{{\bf g}_{k,b}^{(t)}\}_{b=1}^B\big),
    \end{align}	    
    where ${\sf Concatenate}\big(\{{\bf g}_{k,b}^{(t)}\}_{b=1}^B\big)$ is a function that concatenates $B$ sub-vectors $\{{\bf g}_{k,b}^{(t)}\}_{b=1}^B$ into the form of the original vector ${\bf g}_k^{(t)}$.
    The residual vector ${\bf \Delta}_{k}^{(t+1)}$ is stored at device $k$ and then added to the next local gradient vector at iteration $t+1$: 
 	\begin{align}\label{eq:local_grad}
         \bar{\bf g}_{k}^{(t+1)} = \nabla F_{k}^{(t+1)}\big({\bf w}_{t+1}\big)  + {\bf \Delta}_{k}^{(t+1)}.
    \end{align}	  
    The local gradient vector in \eqref{eq:local_grad} is applied as the input of the block sparsification process at iteration $t+1$.

	%To minimize information loss caused by the gradient dropping, each device accumulates the local gradient entries that have not been transmitted until the current communication round.
	%Let ${\bm \Delta}_{k,b}^{(t)} \in \mathbb{R}^{N}$ be the accumulated gradient vector stored for $\nabla f_{k,b}({\bf w}_t;\mathcal{B}_k^{(t)})$. 
	%Then device $k$ sparsifies the sum of the current and the accumulated gradient vectors by setting all but the $S$ largest elements of ${\bm \Delta}_{k,b}^{(t)} + \nabla f_{k,b}({\bf w}_t;\mathcal{B}_k^{(t)})$ to zero, where $S$ is a sparsification level such that $S \ll N$.

	%\subsection{Compress-and-Quantize Process}
	{\bf Dimension Reduction:}
	In the dimension reduction process, each gradient sub-vector is projected onto a lower dimensional space using a sensing matrix ${\bf A} \in \mathbb{R}^{M\times N}$ with $M<N$. 
	%To be specific, each gradient sub-vector ${\bf g}_{k,b}^{(t)}$ is multiplied with a sensing matrix ${\bf A} \in \mathcal{\bf R}^{M\times N}$ with $M<N$, 
	%the dimension of each gradient sub-vector is reduced by random projection onto a lower dimensional space.
	%By denoting $M < N$ as the dimension of the projected space, 
	Then a low-dimensional projection of a gradient sub-vector ${\bf g}_{k,b}^{(t)}$ is represented as 
	\begin{align}\label{eq:compress_block}
        {\bf x}_{k,b}^{(t)}  = \alpha_{k,b}^{(t)} {\bf A} {\bf g}_{k,b}^{(t)},
    \end{align}	
    where $\alpha_{k,b}^{(t)}$ is a scaling factor.
    In this process, the choice of the sensing matrix ${\bf A}$ is critical to successful recovery of local gradients at the PS.
    In CS theory, it is shown that if ${\bf A}$ is an independent and identically distributed (IID) random matrix with $({\bf A})_{m,n} \sim \mathcal{N}(0,1/M)$ for $M = \mathcal{O}\big(S \log (N/S)/\delta_{2S}^2\big)$, there exists $\delta_{2S}\in(0,1)$ such that 
    $(1-\delta_{2S}) \|{\bf g}\|^2 \leq \| {\bf A}{\bf g}\|^2  \leq (1+\delta_{2S}) \|{\bf g}\|^2$
    % \begin{align}\label{eq:RIP}
    %     (1-\delta_{2S}) \|{\bf g}\|^2 \leq \| {\bf A}{\bf g}\|^2  \leq (1+\delta_{2S}) \|{\bf g}\|^2,
    % \end{align}	 
    with high probability, which is called the restricted isometry property (RIP) of order $2S$ \cite{CS:Book}. 
    This condition is known to be sufficient for a variety of algorithms to enable successful recovery of a sparse signal from noisy linear measurements \cite{CS:Book}.
    Motivated by this fact, we set ${\bf A}$ as an IID random matrix with $({\bf A})_{m,n} \sim \mathcal{N}(0,1/M)$.
    %It is well known that for an i.i.d. random matrix ${\bf A}$ with $({\bf A})_{m,n} \sim \mathcal{N}(0,1/M)$, there exists a $\delta_{2S}\in(0,1)$ such that for any $S$-sparse vector ${\bf g}$, the following holds
    %One of the desirable properties to guarantee the feasibility of the reconstruction is a restricted isometry property (RIP) [XX]. 
    %By choosing a matrix ${\bf A}$ as an i.i.d. random matrix with $({\bf A})_{m,n} \sim \mathcal{N}(0,1/M)$, the RIP of order $2S$ is satisfied with high probability provided that $N \geq c S \ln(N/S)$ with a small constant $c$ [XX]. 
    %and ${\bf A} \in \mathbb{R}^{M \times N}$ is a random measurement matrix with $M < N$.
	%By concatenating these $B$ compressed sub-vectors, a compressed gradient vector at device $k$ is obtained as 
	%${\bf x}_k^{(t)} = \big[({\bf x}_{k,1}^{(t)})^{\sf T}, \cdots, ({\bf x}_{k,B}^{(t)})^{\sf T}\big]^{\sf T}$.
	Under this choice, it is expected that $\mathbb{E}\big[ \|{\bf x}_{k,b}^{(t)}\|^2\big] = \big(\alpha_{k,b}^{(t)}\big)^2 \|{\bf g}_{k,b}^{(t)}\|^2$. 
	Based on this fact, we also set the scaling factor as $\alpha_{k,b}^{(t)}={\sqrt{M}}/{\|{\bf g}_{k,b}^{(t)}\|}$ which guarantees that every low-dimensional sub-vector has an equal power of $M$, for all $k,b,t$. 
% 	Considering this choice, we set the scaling factor as $\alpha_{k,b}^{(t)}={\sqrt{M}}/{\|{\bf g}_{k,b}^{(t)}\|}$ so that 
% 	every entry of ${\bf x}_{k,b}^{(t)}$ has a unit variance:
%     \begin{align} 
%         \mathbb{E}\big[|x_{k,b,m}^{(t)}|^2\big] 
%         &= \big(\alpha_{k,b}^{(t)}\big)^2 \mathbb{E}\big[|{\bf a}_{{\rm row},m}{\bf g}_{k,b}^{(t)}|^2\big] 
%         % \nonumber \\
%         % &= \frac{M}{\|{\bf g}_{k,b}^{(t)}\|^2} ({\bf g}_{k,b}^{(t)})^{\sf T}\mathbb{E}\big[{\bf a}_{{\rm row},m}{\bf a}_{{\rm row},m}^{\sf H}\big]{\bf g}_{k,b}^{(t)} \nonumber \\
%         &= \frac{M}{\|{\bf g}_{k,b}^{(t)}\|^2} \times \frac{\|{\bf g}_{k,b}^{(t)}\|^2}{M} = 1, \label{eq:x_var}
%     \end{align}	
%     where $x_{k,b,m}^{(t)}$ is the $m$-th entry of ${\bf x}_{k,b}^{(t)}$, ${\bf a}_{{\rm row},m}$ is the $m$-th row of ${\bf A}$, and the equality (a) holds because each element of ${\bf A}$ is independently drawn from $\mathcal{N}(0,1/M)$.
    This property will be utilized in the design of a scalar quantizer in the subsequent quantization process.

	{\bf Quantization:}
	In the quantization process, each entry of a low-dimensional sub-vector is quantized by using a $Q$-bit scalar quantizer. 
	Let ${\sf Q}: \mathbb{R} \rightarrow \mathcal{Q}\triangleq \{q_1,\ldots,q_{2^Q}\}$ be a $Q$-bit scalar quantizer that maps a real value input to the nearest point in $\mathcal{Q}$, i.e., ${\sf Q}(x) = q_i$ if $x \in (\tau_{i-1},\tau_i]$,
%    \begin{align}\label{eq:quantizer}
%        {\sf Q}(x) = q_i~~\text{if}~~ x \in (\tau_{i-1},\tau_i],
%        % = \begin{cases}
%        %     q_1, & x< \tau_1, \\
%        %     q_i, & \tau_{i-1} \leq x < \tau_i,~i\in\{2,\ldots,2^Q-1\},\\
%        %     q_{2^Q}, & \tau_{2^Q-1} < x,
%        % \end{cases}
%    \end{align}	
    where $q_i$ is the $i$-th quantizer output, and $\tau_i$ is the $i$-th quantizer threshold with $\tau_0 <  \ldots < \tau_{2^Q}$ with $\tau_0 = -\infty$ and $\tau_{2^Q} = \infty$.  
    Then the quantized sub-vector at device $k$ is obtained as 
    \begin{align}\label{eq:quantize_block}
        {\bf q}_{k,b}^{(t)} = {\sf Q}\big({\bf x}_{k,b}^{(t)}\big).
    \end{align}	
    Since the accuracy of gradient reconstruction at the PS is closely related to a quantization error, given by ${\sf Q}\big({\bf x}_{k,b}^{(t)}\big) -{ \bf x}_{k,b}^{(t)}$, we also optimize the design of the scalar quantizer to minimize the MSE of the quantizer output by leveraging the Lloyd-Max algorithm \cite{LloydMax}. 
    %To maximize the accuracy of gradient reconstruction at the PS, we employ the optimal scalar  quantizer that minimizes a quantization error ${\sf Q}\big({\bf x}_{k,b}^{(t)}\big) -{ \bf x}_{k,b}^{(t)}$. 
    To this end, the knowledge of the distribution of ${\bf x}_{k,b}^{(t)}$ is required at the devices, which is challenging in federated learning due to the difficulty in characterizing the distribution of local gradients.
    To circumvent this challenge, instead of characterizing the exact distribution, we model each local gradient sub-vector as an IID random vector by using an approximate distribution (e.g., a Bernoulli Gaussian-mixture distribution), as will be justified in Sec.~\ref{Sec:QGAMP}.
    Under this approximate model, the projection in \eqref{eq:compress_block} with $({\bf A})_{m,n} \sim \mathcal{N}(0,1/M)$ implies that each entry of ${\bf x}_{k,b}^{(t)}$ behaves like a zero-mean Gaussian random variable for large $N$ by the central limit theorem.
    Meanwhile, the choice of $\alpha_{k,b}^{(t)}={\sqrt{M}}/{\|{\bf g}_{k,b}^{(t)}\|}$ implies that $\mathbb{E}\big[ \|{\bf x}_{k,b}^{(t)}\|^2\big] = M$.
    Therefore, every entry of ${\bf x}_{k,b}^{(t)}$ can be effectively modeled by a Gaussian random variable with zero mean and unit variance. 
    Motivated by this fact, we employ the Lloyd-Max scalar quantizer optimized for the distribution of $\mathcal{N}(0,1)$.
    A key benefit of our optimization is that the design of the scalar quantizer does not depend on the indexes $k,b,t$; thereby, the optimal scalar quantizer can be shared by all the devices and the PS in prior without explicit information exchange. 

	\begin{algorithm}
		\caption{Federated Learning via Quantized Compressed Sensing (FedQCS)}\label{alg:FedQCS}
		{\small
		{\begin{algorithmic}[1]
			\REQUIRE Initial parameter vector ${\bf w}_1$, ${\bf \Delta}_{k}^{(1)}={\bf 0}_{\bar{N}}$, $\{\mathcal{N}_b\}_{b=1}^B$, $\{\mathcal{K}_g\}_{g=1}^G$ \\
            \ENSURE Optimized parameter vector ${\bf w}_T$
			\FOR {$t=1$ to $T$}
			    \STATE \!\!\!{\em At the wireless devices:}
				\FOR {Each device $k\in\mathcal{K}$}
    				\STATE $\bar{\bf g}_{k}^{(t)} = \nabla F_{k}^{(t)}\big({\bf w}_t\big)  + {\bf \Delta}_{k}^{(t)}$.
    				\STATE $\{{\bf g}_{k,b}^{(t)}\}_{b=1}^B = {\sf BlockSparse}\big(\bar{\bf g}_{k}^{(t)}\big)$.
                    %\STATE ${\bf x}_{k,b}^{(t)} = {\bf A}{\bf g}_{k,b}^{(t)}$, $\forall b\in\{1,\ldots,B\}$.
                    \STATE ${\bf \Delta}_{k}^{(t+1)} = \bar{\bf g}_{k}^{(t)} - {\sf Concatenate}\big(\{{\bf g}_{k,b}^{(t)}\}_{b=1}^B\big)$.
                    \STATE ${\bf q}_{k,b}^{(t)} = {\sf Q}\big({\bf x}_{k,b}^{(t)}\big)$ where ${\bf x}_{k,b}^{(t)}=\alpha_{k,b}^{(t)}{\bf A}{\bf g}_{k,b}^{(t)}$, $\forall b$.
                    \STATE Push $\big\{{\bf q}_{k,b}^{(t)}, \alpha_{k,b}^{(t)}\big\}_{b=1}^B$ to the parameter server.
		        \ENDFOR
    			\STATE \!\!\!{\em At the parameter server:}
    			\IF{Estimate-and-aggregate strategy}
			        \STATE $\hat{\bf g}_{k,b}^{(t)} = {\sf QEMGAMP}\big({\bf q}_{k,b}^{(t)}, \alpha_{k,b}^{(t)},{\bf A}\big)$ from {\bf Procedure~\ref{alg:QEMGAMP}}, $\forall k, b$.
			        \STATE $\hat{\bf g}_{k}^{(t)} = {\sf Concatenate}\big(\{\hat{\bf g}_{k,b}^{(t)}\}_{b=1}^B\big)$, $\forall k$.
    			    \STATE $\hat{\bf g}_{\mathcal{K}}^{(t)} = \sum_{k=1}^K \rho_k^{(t)} \hat{\bf g}_{k}^{(t)}$.
			    \ELSIF{Aggregate-and-estimate strategy}
			        \STATE $\tilde{\bf q}_{\mathcal{K}_g,b}^{(t)} = \sum_{k\in\mathcal{K}_g} \big\{\rho_k^{(t)}/(\gamma_Q\alpha_{k,b}^{(t)}) \big\} {\bf q}_{k,b}^{(t)}$, $\forall g, b$.
			        \STATE $\nu_{g,b}^{(t)} = {(\psi_Q - \gamma_{Q}^2)}/{\gamma_{Q}^2} \sum_{k\in\mathcal{K}_g} \big\{{\rho_k^{(t)}}/{\alpha_{k,b}^{(t)}}\big\}^2$, $\forall g, b$.
			        \STATE $\hat{\bf g}_{\mathcal{K}_g,b}^{(t)} = {\sf EMGAMP}\big(\tilde{\bf q}_{\mathcal{K}_g,b}^{(t)},\nu_{g,b}^{(t)},{\bf A}\big)$, $\forall g, b$.
			        %\STATE Compute $\tilde{\bf q}_{\mathcal{K}_g,b}^{(t)}$ and $\nu_{g,b}^{(t)}$ from \eqref{eq:joint_recovery} and \eqref{eq:distortion_var}, respectively.
    			    %\STATE $\hat{\bf g}_{\mathcal{K}_g,b}^{(t)} = {\sf BussEMGAMP}\big(\{{\bf q}_{k,b}^{(t)}, \alpha_{k,b}^{(t)}\}_{k\in\mathcal{K}_g}\big)$ from {\bf Procedure~\ref{alg:BussEMGAMP}}, $\forall g, b$.
    			    %\STATE Replace Step 7 and Step 8 of {\bf Procedure~\ref{alg:QEMGAMP}} with \eqref{eq:Buss_x_post} and \eqref{eq:Buss_nu_post}, respectively, where $\nu_{d} = \nu_{g,b}^{(t)}$, $\forall g, b$.
    			    %\STATE $\hat{\bf g}_{\mathcal{K}_g,b}^{(t)} = {\sf QEMGAMP}\big(\tilde{\bf q}_{\mathcal{K}_g,b}^{(t)},1\big)$ from {\bf Procedure~\ref{alg:QEMGAMP}}, $\forall g, b$.
    			    \STATE $\hat{\bf g}_{\mathcal{K}_g}^{(t)} = {\sf Concatenate}\big(\{\hat{\bf g}_{\mathcal{K}_g,b}^{(t)}\}_{b=1}^B\big)$, $\forall g$.
    			    \STATE $\hat{\bf g}_{\mathcal{K}}^{(t)} = \sum_{g=1}^G \hat{\bf g}_{\mathcal{K}_g}^{(t)}$.
    			\ENDIF
    			%\STATE Compute ${\bf w}_{t+1}$ based on 
    			\STATE ${\bf w}_{t+1} = {\bf w}_t - \eta_t \hat{\bf g}_{\mathcal{K}}^{(t)}$.
    			%\STATE ${\bf w}_{t+1} = {\sf Update}({\bf w}_t,\hat{\bf g}_{\rm glo}^{(t)})$
    			\STATE Broadcast ${\bf w}_{t+1}$ to the wireless devices.
			\ENDFOR
		\end{algorithmic}}}
	\end{algorithm}

    \subsection{Communication Overhead of BQCS Compression}
    When employing the proposed BQCS compression, each device $k$ needs to convey the information of $\big\{ {\bf q}_{k,b}^{(t)}, \alpha_{k,b}^{(t)}\big\}_{b=1}^B$ to the PS. 
    %FedQCS with the proposed compression strategy is summarized in {\bf Procedure~\ref{alg:FedQCS}}.
    %This clearly demonstrates that the communication overhead per each device is $QMB$ bits because each element of ${\bf q}_{k,b}^{(t)}$ can be represented using $Q$ bits, while 
    Note that $Q$ bits are required to transmit each entry of ${\bf q}_{k,b}^{(t)}$, while $32$ bis are required to transmit $\alpha_{k,b}^{(t)}$ using floating-point representation.
    Since $32$-bit overhead is negligible compared to $QM$ bits, communication overhead of the proposed compression becomes $\frac{QMB}{NB}= \frac{Q}{R}$ bits per gradient entry, where $R\triangleq \frac{N}{M} >1$ is a dimension reduction ratio. 
    %\footnote{When employing floating-point representation, $32$ bits are required to transmit the scaling factor $\alpha_{k,b}^{(t)}$. This overhead is negligible compared to $QM$ bits because $M\gg 1$ in general federated learning settings.}.
    %Therefore, communication overhead of the proposed compression is $QMB$ bits per device as 
	%In other words, communication overhead per gradient entry is only $\frac{QMB}{NB}= \frac{Q}{R}$ bits, where $R\triangleq \frac{N}{M} >1$ is a dimension reduction ratio. 
	This fact clearly reveals that the communication overhead of the BQCS compression is adjustable by changing the number of quantization bits, $Q$, and the dimension reduction ratio, $R$. 
	A more important observation is that by choosing $Q<R$, the communication overhead of our compression can be made even less than one bit per gradient entry.
	Thanks to this feature, our strategy achieves a higher compression ratio than the state-of-the-art one-bit gradient compression \cite{SignSGD}.
	%This implies that the BQCS compression with $Q<R$ requires less overhead than one-bit gradient compression which has been widely considered in the literature (e.g., signSGD in \cite{SignSGD}). 
    % Therefore, the communication overhead of FedQCS is given by $\frac{QMB}{NB}= \frac{Q}{R}$ bits per gradient entry, where $R\triangleq \frac{N}{M} >1$ is a compression ratio.
    % A key implication of the above fact is that the communication overhead of FedQCS can be adjusted by controlling the number of quantization bits, $Q$, and the compression ratio, $R$. 
    % Another important implication is that by choosing $Q<R$, the communication overhead of FedQCS can be made even less than that of one-bit quantization which is well-known as a gradient compression technique for federated learning (e.g., signSGD in \cite{SignSGD}).
    In Sec.~\ref{Sec:Simul}, we will also demonstrate that the BQCS compression enables more accurate reconstruction of the global gradient vector at the PS compared to the one-bit gradient compression.

	\vspace{3mm}
    {\bf Remark 1 (Comparison to Existing QCS-based Gradient Compression):}
    We highlight the major differences between the BQCS compression and the existing QCS-based gradient compression methods in \cite{DitherQCS,QCS1,QCS2}. 
    %A federated learning framework based on QCS has also been studied in \cite{DitherQCS,QCS1,QCS2}. 
    The compression method in \cite{DitherQCS} adopts dithered uniform quantization combined with dimension reduction.
    A key advantage of the dithered quantization is that the quantized signal can be transformed into the sum of a quantizer input signal and an independent quantization noise when the uniform quantizer is adopted with random dither signals \cite{DitherTheory}. 
    This advantage, however, is attained at the cost of additional communication overhead because the information of the dither signal (with dimension $\bar{N}$) should be separately conveyed to the PS for gradient reconstruction.
    In addition, the use of the uniform quantizer leads to larger quantization error compared to the optimal quantizer used in our BQCS compression. 
    %In addition, the dithered quantization operates with the uniform quantizer which performs worse than the optimal quantizer considered in our framework. 
    The compression method in \cite{QCS1,QCS2} adopts one-bit scalar quantization combined with dimension reduction, which is also known as one-bit compressed sensing.
    In this method, the number of quantization bits is fixed to only one (i.e., $Q=1$); thereby, communication overhead of this method is less flexible than that of the BSDQ compression.
    In Sec.~\ref{Sec:Simul}, we will also demonstrate that our BQCS compression enables more accurate gradient reconstruction at the PS compared to the existing compression methods under the same communication overhead.

    \section{Gradient Reconstruction Strategies of FedQCS}\label{Sec:Recovery}
    One of the primary goals at the PS is to accurately reconstruct the global gradient vector in \eqref{eq:global_grad} from the local gradient vectors sent by wireless devices, in order to optimize the parameter vector ${\bf w}$. 
	Unfortunately, the use of the BQCS compression in Sec.~\ref{Sec:Comp} brings a new challenge in realizing accurate gradient reconstruction at the PS because the local gradient vectors sent by the device are not only projected onto a low-dimensional space as done in CS, but also nonlinearly distorted by scalar quantization. 
    %One of the primary goals at the PS is to minimize the mismatch between the global gradient vector in \eqref{eq:global_grad} and its reconstruction from the compressed local gradient vectors sent by wireless devices. 
    %Unfortunately, accurate estimation of the local gradients at PS is challenging because the local gradient sent by the device are nonlinearly distorted by sparsification and quantization. 
    %Unfortunately, achieving this goal is challenging in FedQCS because each local gradient vector is not only compressed but also quantized in order to reduce the communication overhead of federated learning.
    %Although the QCS-based compression scheme in Sec.~\ref{Sec:Comp} significantly reduces the communication overhead of federated learning, this scheme brings a new challenge when reconstructing a global gradient vector at the PS. 
    In this section, we tackle this challenge by presenting two gradient reconstruction strategies, referred to as \textit{estimate-and-aggregate} and {\em aggregate-and-estimate}, which enable accurate aggregation of the local gradients from the compressed signals. 

	\subsection{Estimate-and-Aggregate Strategy}\label{Sec:QGAMP}
	The key idea of the estimate-and-aggregate strategy is to first estimate each local gradient sub-vector from its compressed signal, then aggregate the estimated sub-vector to reconstruct a global gradient vector. 
	In this strategy, the problem of estimating a local gradient sub-vector, ${\bf g}_{k,b}^{(t)}$, from its compressed signal, ${\bf q}_{k,b}^{(t)} = {\sf Q}\big(\alpha_{k,b}^{(t)}{\bf A}{\bf g}_{k,b}^{(t)}\big)$, is nothing but a QCS recovery problem. 
	%to convert a global gradient reconstruction problem into $KB$ parallel QCS recovery problems, each of which aims at reconstructing a local gradient sub-vector, ${\bf g}_{k,b}^{(t)}$, from its \textit{quantized} linear observation, ${\bf q}_{k,b}^{(t)} = {\sf Q}\big(\alpha_{k,b}^{(t)}{\bf A}{\bf g}_{k,b}^{(t)}\big)$.
	%we employ a variation of the EM-GAMP algorithm in \cite{EMGAMP} which provides an approximate MMSE solution of the QCS recovery problem on the basis of Bayesian inference.
% 	The fundamental idea of the parallel recovery strategy is to convert a global gradient reconstruction problem into $KB$ parallel QCS recovery problems, each of which estimates each device's local gradient sub-vector, ${\bf g}_{k,b}^{(t)}$, from the associated quantized sub-vector, ${\bf q}_{k,b}^{(t)}$. 
	There are two underlying challenges to solve this problem: (i) finding the optimal solution of a QCS recovery problem in terms of minimizing the reconstruction error is still an open problem \cite{QEMGAMP,QGAMP}, and (ii) the distribution of the local gradient sub-vectors is generally unknown at the PS which prevents the PS from directly applying a Bayesian inference approach. 
	%the distribution of the local gradient sub-vectors is unknown at the PS, which makes it challenging to find the optimal solution of the QCS recovery problem in terms of minimizing reconstruction error.
 	To circumvent these challenges, we employ a quantized EM-GAMP (Q-EM-GAMP) algorithm in \cite{QEMGAMP} which iteratively computes an {\em approximate} MMSE solution of a QCS recovery problem while learning the distribution of an unknown signal via the EM principle. 
 	%which iteratively computes an {\em approximate} MMSE solution of a QCS recovery problem on the basis  while learning the distribution of an unknown signal via the EM principle. 
	%We refer to this algorithm as a {\em quantized-EM-GAMP (Q-EM-GAMP)} algorithm because it modifies the original EM-GAMP algorithm by taking into account the effect of the scalar quantization.
	%A similar variant of the EM-GAMP algorithm is also studied in \cite{QEMGAMP}.

	%A key ingredient of the Q-EM-GAMP algorithm is statistical modeling of an unknown signal which enables the use of the Bayesian inference for reconstructing this signal. 
	%To employ this algorithm to solve our QCS recovery problem, we model each local gradient sub-vector as an i.i.d. random vector by introducing a proper \textit{probabilistic} distribution. 
	%Particularly, inspired by the sparse property of the local gradient sub-vector, we adopt a Bernoulli Gaussian-mixture distribution for such distribution, which is well known for its suitability and generality for modeling a sparse random vector \cite{EMGAMP,QEMGAMP}.  
	To employ the Q-EM-GAMP algorithm to solve our QCS recovery problem, each local gradient sub-vector needs to be modeled as an IID random vector with a proper distribution. 
	Inspired by the sparse property as well as the arbitrary random nature of the local gradient sub-vector, we model each sub-vector using a Bernoulli Gaussian-mixture distribution which is well known for its suitability and generality for modeling a sparse random vector \cite{EMGAMP,QEMGAMP}.   
	Note that the probability density function of the Bernoulli Gaussian-mixture distribution with parameter ${\bm \theta} = (\lambda_0,\{\lambda_l,\mu_l,\phi_l\}_{l=1}^L)$ is given by 
% 	In particular, each gradient sub-vector is modeled as an i.i.d. random vector whose entry follows a Bernoulli Gaussian-mixture distribution with parameter ${\bm \theta} = (\lambda_0,\{\lambda_l,\mu_l,\phi_l\}_{l=1}^L)$ and probability density function:
	\begin{align}\label{eq:BernoulliGM}
        \mathcal{BG}(g;\boldsymbol{\theta}) = \lambda_0\delta(g) + \sum_{l=1}^L \frac{\lambda_l}{\sqrt{2\pi\phi_l}}{\rm exp}\bigg(-\frac{(g-\mu_l)^2}{2\phi_l}\bigg).
    \end{align}	
	%The Bernoulli Gaussian-mixture model in \eqref{eq:BernoulliGM} can be interpreted as follows: First of all, 
	As can be seen in \eqref{eq:BernoulliGM}, the sparse property of the local gradient sub-vector is captured by the Bernoulli distribution with parameter $\lambda_0$. 	
	Meanwhile, the arbitrary random nature of non-zero entries can be effectively approximated by a Gaussian-mixture distribution with $L$ components, where the mean and the variance of the $l$-th component are denoted by $\mu_l$ and $\phi_l$, respectively. 
	%Utilizing this distribution, in the Q-EM-GAMP algorithm, each sub-vector ${\bf g}_{k,b}^{(t)}$, is modeled as an i.i.d. random vector whose entry follows a Bernoulli Gaussian-mixture distribution with  parameter ${\bm \theta}_{k,b}^{(t)}$. 
	%How to determine a proper parameter set ${\bm \theta}$ for each gradient sub-vector will be discussed in the sequel.
 	Assuming IID Bernoulli Gaussian-mixture prior, the Q-EM-GAMP algorithm computes an approximate MMSE estimate of ${\bf g}_{k,b}^{(t)}$ from ${\bf q}_{k,b}^{(t)}$ by iterating the following two steps:
	(i) perform the GAMP algorithm to compute the approximate MMSE estimate of $\mathbf{g}_{k,b}^{(t)}$ from $\mathbf{q}_{k,b}^{(t)}$ by assuming that each entry of $\mathbf{g}_{k,b}^{(t)}$ follows the Bernoulli Gaussian-mixture prior with parameter ${\bm \theta}_{k,b}^{(t)}$; 
	(ii) update the parameter ${\bm \theta}_{k,b}^{(t)}$ of the Bernoulli Gaussian-mixture model based on the EM principle. 
	The Q-EM-GAMP algorithm utilized in our parallel recovery strategy is summarized in {\bf Procedure~\ref{alg:QEMGAMP}}, where we omit the indexes $k$, $b$, and $t$, for the sake of brevity.

	\begin{algorithm}
		\caption{The Q-EM-GAMP Algorithm}\label{alg:QEMGAMP}
		{\small
		{\begin{algorithmic}[1]
			\REQUIRE $\mathbf{q} \in \mathcal{Q}^M$,  %$\tilde{\mathbf{g}}^{\rm ini} \in \mathbb{R}^N$,  %$\boldsymbol{\nu}_{\mathbf{g}}^{\rm ini} \in \mathbb{R}^N$
			$\alpha \in \mathbb{R}$, $\mathbf{A} \in \mathbb{R}^{M \times N}$
            \ENSURE $\hat{\mathbf{g}} \in \mathbb{R}^N$
            \STATE Initialize $\hat{g}_n \sim \mathcal{N}(0,\frac{M}{N\alpha^2})$, $\nu_{g_n} = \frac{M}{N\alpha^2}$, and $\boldsymbol{\theta} = (\lambda_0,\{\lambda_l,\mu_l,\phi_l\}_{l=1}^L)$.
            \STATE Set $\hat{s}_m=0$ and $\tilde{a}_{m,n}=\alpha(\mathbf{A})_{m,n}$, $\forall m,n$.
			\FOR {$i=1$ to $I_{\rm GAMP}$}
			    \STATE $\hat{g}^{\rm old}_n=\hat{g}_n$, $\forall n$.
			    \STATE $\nu_{p_m }= \sum_{n=1}^N |\tilde{a}_{m,n}|^2\nu_{g_n}$, $\forall m$.
			    \STATE $\hat{p}_m = \sum_{n=1}^N \tilde{a}_{m,n}\hat{g}_n - \nu_{p_m}\hat{s}_m$, $\forall m$.
			 %   \IF {$Q < \infty$}
			        \STATE $\hat{x}_m^{\rm post} = \mathbb{E}[x_m|q_m, \hat{p}_m, \nu_{p_m }]$ from \eqref{eq:tilde_x_m_post}, $\forall m$. 
			        \STATE $\nu_{x_m}^{\rm post} = {\rm Var}[x_m|q_m, \hat{p}_m, \nu_{p_m }]$ from \eqref{eq:tilde_nu_x_m_post}, $\forall m$.
		  %      \ELSE
			 %       \STATE $\hat{x}_m^{\rm post} = (\hat{p}_m\nu_{d}+q_m\nu_{p_m})/(\nu_{p_m}+\nu_{d})$, $\forall m$.
			 %       \STATE $\nu_{x_m}^{\rm post} = (1/\nu_{p_m}+1/\nu_{d})^{-1}$, $\forall m$.
			 %   \ENDIF
			    \STATE $\hat{s}_m = (\hat{x}_m^{\rm post} - \hat{p}_m)/\nu_{p_m}$, $\forall m$. 
			    \STATE $\nu_{s_m} = (1-\nu_{x_m}^{\rm post}/\nu_{p_m})/\nu_{p_m}$, $\forall m$.
			    \STATE $\hat{r}_n = \hat{g}_n + \nu_{r_n}\sum_{m=1}^{M}\tilde{a}_{m,n}\hat{s}_m$, $\forall n$. 
			    \STATE $\nu_{r_n} = (\sum_{m=1}^{M}|\tilde{a}_{m,n}|^2 \nu_{s_m})^{-1}$, $\forall n$.
			    \STATE $\hat{g}_n = \sum_{l=1}^L \lambda_{n,l}^{\prime}\mu_{n,l}^{\prime}$, $\forall n$.
                \STATE $\nu_{g_n} = \sum_{l=1}^L \lambda_{n,l}^{\prime}(\phi_{n,l}^{\prime} + (\mu_{n,l}^{\prime})^2) - (\hat{g}_n)^2$, $\forall n$. 
			    \STATE ${\bm \theta} \leftarrow \big(\lambda_0^{\prime \prime}, \{\lambda_l^{\prime \prime},\mu_l^{\prime \prime},\phi_l^{\prime \prime} \}_{l=1}^L\big)$ from \eqref{eq:param_GM}.
			    \STATE Break if $\sum_{n=1}^N (\hat{g}^{\rm old}_n-\hat{g}_n)^2 < \tau_{\rm GAMP}\sum_{n=1}^N (\hat{g}^{\rm old}_n)^2$.
			    %\STATE $\hat{g}_n^{\rm old} = \hat{g}_n$, $\forall n$.
			\ENDFOR
			\STATE $\hat{\mathbf{g}} = [\hat{g}_1, \cdots ,\hat{g}_N]^{\rm T}$.
		\end{algorithmic}}}
	\end{algorithm}	
	
    %We also elaborate on the details of the Q-EM-GAMP algorithm employed in our parallel recovery strategy. The fundamental of this algorithm is to iteratively compute an {\em approximate} MMSE solution of a QCS recovery problem while learning the distribution of an unknown signal via the EM principle.

	The major steps in {\bf Procedure~\ref{alg:QEMGAMP}} are elaborated below.
	In Steps 5 and 6, the prior mean $\hat{p}_m$ and variance $\nu_{p_m}$ of $x_m$ are estimated. 
	In Steps 7 and 8, the posterior mean and variance of $x_m$ are computed under the assumption of $x_m \sim \mathcal{N}(\hat{p}_m, \nu_{p_m })$, given by
	\begin{align}
        \hat{x}_m^{\rm post} &= \int_{-\infty}^{\infty}x_m p(x_m | q_m)\,{\rm d}x_m = \hat{p}_m + \nu_{p_m}\frac{p'(q_m)}{p(q_m)}, \label{eq:tilde_x_m_post} \\
        \nu_{x_m}^{\rm post} &= \int_{-\infty}^{\infty}x_m^2 p(x_m | q_m)\,{\rm d}x_m - (x_m^{\rm post})^2
        = \nu_{p_m}^2\bigg\{ \frac{p''(q_m)}{p(q_m)} - \bigg(\frac{p'(q_m)}{p(q_m)}\bigg)^{\!2}\bigg\} + \nu_{p_m}, \label{eq:tilde_nu_x_m_post}
    \end{align}
    where 
	\begin{align}\label{eq:Bayes_denomi_deriv}
	    p(q_m) &= 
        %\int_{-\infty}^{\infty}p(q_m | x_m)p(x_m)\,{\rm d}x_m =
        Q\Bigg(\frac{\tau_{i-1}-\hat{p}_m}{\sqrt{\nu_{p_m}}}\Bigg) - Q\Bigg(\frac{\tau_{i}-\hat{p}_m}{\sqrt{\nu_{p_m}}}\Bigg), \\
        p'(q_m) 
        &= \frac{1}{\sqrt{\nu_{p_m}}}\Bigg[\phi\Bigg(\frac{\tau_{i-1}-\hat{p}_m}{\sqrt{\nu_{p_m}}}\Bigg) - \phi\Bigg(\frac{\tau_{i}-\hat{p}_m}{\sqrt{\nu_{p_m}}}\Bigg)\Bigg], \\
        p''(q_m)  
        %&= \int_{\tau_{i-1}}^{\infty}\bigg(\frac{x_m-\hat{p}_m}{\nu_{p_m}} \bigg)^2 \mathcal{N}(\hat{p}_m, \nu_{p_m})\,{\rm d}x_m - \int_{\tau_{i}}^{\infty}\bigg(\frac{x_m-\hat{p}_m}{\nu_{p_m}} \bigg)^2 \mathcal{N}(\hat{p}_m, \nu_{p_m})\,{\rm d}x_m         \nonumber \\
        %&~~~- \int_{\tau_{i-1}}^{\infty} \frac{1}{\nu_{p_m}}\mathcal{N}(\hat{p}_m, \nu_{p_m})\,{\rm d}x_m + \int_{\tau_{i}}^{\infty}\frac{1}{\nu_{p_m}}\mathcal{N}(\hat{p}_m, \nu_{p_m})\,{\rm d}x_m    \nonumber \\
        &= \frac{1}{\nu_{p_m}} \Bigg[ \phi\Bigg(\frac{\tau_{i-1}-\hat{p}_m}{\sqrt{\nu_{p_m}}}\Bigg)\Bigg(\frac{\tau_{i-1}-\hat{p}_m}{\sqrt{\nu_{p_m}}}\Bigg) - \phi\Bigg(\frac{\tau_{i}-\hat{p}_m}{\sqrt{\nu_{p_m}}}\Bigg)\Bigg(\frac{\tau_{i}-\hat{p}_m}{\sqrt{\nu_{p_m}}}\Bigg)
        \Bigg],
    \end{align}
    with $Q(x)=\int_{x}^{\infty}\frac{1}{\sqrt{2\pi}}e^{-\frac{u^2}{2}}{\rm d}u$ and $\phi(x)=\frac{1}{\sqrt{2\pi}}e^{-\frac{x^2}{2}}$.
    In Steps 11 and 12, $\hat{r}_n$ represents the observation of $g_n$ under zero-mean Gaussian noise, while $\nu_{r_n}$ represents the variance of the noise.
    %the estimate of $g_n$ and the corresponding error variance are determined.
	In Steps 13 and 14, the posterior mean and variance of $g_n$ are computed under the assumptions of $ g_n \sim \mathcal{BG}({\bm \theta})$ and $\hat{r}_n = g_n + \xi_n$ with $\xi \sim \mathcal{N}(0,\nu_{r_n})$ as derived in \cite{EMGAMP}, where $\beta_{n,0} = \lambda_0\mathcal{N}(0;\hat{r}_n,\nu_{r_n})$, $\beta_{n,l} = \lambda_l\mathcal{N}(\hat{r}_n;\mu_l,\nu_{r_n}+\phi_l)$, $\lambda_{n,0}^{\prime} = {\beta_{n,0}}/({\beta_{n,0}+\sum_{i=1}^L \beta_{n,i}})$, $\lambda_{n,l}^{\prime} = {\beta_{n,l}}/({\beta_{n,0}+\sum_{i=1}^L \beta_{n,i}})$, $\mu_{n,l}^{\prime} = ({\hat{r}_n\phi_l + \mu_l \nu_{r_n}})/({\nu_{r_n}+\phi_l})$, $\phi_{n,l}^{\prime} = {\nu_{{r}_n}\phi_l}/({\nu_{r_n}+\phi_l})$,
	and $\mathcal{N}(x;\mu_x,\nu_x) = \frac{1}{{\sqrt{2\pi \nu_x}}}e^{-\frac{(x-\mu_x)^2}{2\nu_x}}$.
	%\begin{align}
    	%\lambda_{n,0}^{\prime} = \frac{\beta_{n,0}}{\beta_{n,0}+\sum_{i=1}^L \beta_{n,i}},\quad \lambda_{n,l}^{\prime} = \frac{\beta_{n,l}}{\beta_{n,0}+\sum_{i=1}^L \beta_{n,i}},\quad
    	%\mu_{n,l}^{\prime} = \frac{\hat{r}_n\phi_l + \mu_l \nu_{r_n}}{\nu_{r_n}+\phi_l},\quad
    	%\phi_{n,l}^{\prime} = \frac{\nu_{r_n}\phi_l}{\nu_{r_n}+\phi_l}, \nonumber
	%\end{align}
	%for all $ l \in \{ 1,\ldots,L \} $.	
	In Step 15, the parameters of the Bernoulli Gaussian-mixture model in \eqref{eq:BernoulliGM} are computed based on the EM principle as derived in \cite{EMGAMP}, where $\lambda_0^{\prime\prime} = \frac{1}{N}\sum_{n=1}^{N}\lambda_{n,0}^{\prime}$,
	\begin{align}\label{eq:param_GM}
    	\lambda_l^{\prime\prime} \approx \frac{1}{N}\sum_{n=1}^N \lambda_{n,l}^{\prime},\quad 
    	\mu_l^{\prime\prime} \approx \frac{\sum_{n=1}^N \lambda_{n,l}^{\prime}\mu_{n,l}^{\prime}}{\sum_{n=1}^N\lambda_{n,l}^{\prime}},\quad
    	\phi_{n,l}^{\prime\prime} \approx \frac{\sum_{n=1}^N \lambda_{n,l}^{\prime} \big\{ (\mu_l - \mu_{n,l}^{\prime})^2 + \phi_{n,l}^{\prime} \big\}}{\sum_{n=1}^N \lambda_{n,l}^{\prime}},
	\end{align}
	for $ l \in \{ 1,\ldots,L \}$.

% 	\vspace{3mm}
% 	{\bf Remark 3 (Comparison to Existing GAMP Algorithms):} 
% 	The proposed Q-EM-GAMP algorithm generalizes both the EM-GAMP algorithm in \cite{EMGAMP} and the quantized GAMP algorithm in [XX] in different directions. 
% 	On the one hand, the EM-GAMP algorithm in \cite{EMGAMP} estimates a sparse vector from an {\em unquantized} linear observation, while approximating the prior distribution of the sparse vector using the EM principle operating with the Bernoulli Gaussian-mixture model.
% 	The Q-EM-GAMP algorithm extends this algorithm by considering a {\em quantized} linear observation obtained from any scalar quantizer in \eqref{eq:quantizer}.
% 	On the other hand, the quantized GAMP algorithm in [XX] estimates an input vector with a {\em known} prior distribution from a quantized linear observation. 
% 	The Q-EM-GAMP algorithm also extends this algorithm by considering the input vector with an {\em unknown} and {\em sparse} prior distribution.  
% 	Thanks to the general applicability of the Q-EM-GAMP algorithm, it becomes a powerful means of solving a general QCS recovery problem, while overcoming the limitations of both the EM-GAMP algorithm and the quantized GAMP algorithm.  

	After computing the estimates of the local gradient sub-vectors, we aggregate these estimates to reconstruct the global gradient vector in \eqref{eq:global_grad}.
	Let $\hat{\bf g}_{k,b}^{(t)}$ be the estimate of ${\bf g}_{k,b}^{(t)}$ computed by applying the Q-EM-GAMP algorithm to ${\bf q}_{k,b}^{(t)}$. 
	Then the global gradient vector $\hat{\bf g}_{\mathcal{K}}^{(t)}$ is obtained as
	\begin{align}
        \hat{\bf g}_{\mathcal{K}}^{(t)} = \sum_{k=1}^K \rho_k^{(t)} \hat{\bf g}_{k}^{(t)},
    \end{align}	
    where $\hat{\bf g}_{k}^{(t)} = {\sf Concatenate}\big(\{\hat{\bf g}_{k,b}^{(t)}\}_{b=1}^B\big)$.
	The overall gradient reconstruction process of our estimate-and-aggregate strategy is summarized in Steps 12--14 of {\bf Procedure~\ref{alg:FedQCS}}.

	\begin{figure*}
    	\centering
		{\epsfig{file=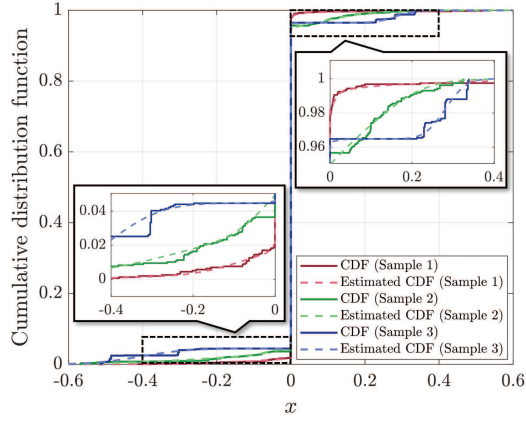, width=7.6cm}} \vspace{-5mm}
    	\caption{Comparison between the empirical CDF of a local gradient sub-vector and an estimated CDF using a Bernoulli Gaussian-mixture distribution.}  \vspace{-5mm}
    	\label{fig:Bernoulli}
    \end{figure*}	
	
	\vspace{3mm}
	{\bf Validation for IID Bernoulli Gaussian-mixture prior:} 	  
	We verify the tightness of the statistical model utilized in the Q-EM-GAMP algorithm using a simple numerical example.
    In this example, we consider an image classification task using the MNIST dataset when $(R,Q)=(3,3)$ and $T=1$. Further details of the simulation are described in Sec.~\ref{Sec:Simul}.
    %We draw the samples of ${\bf g}_{k,b}^{(t)}$ and ${\bf x}_{k,b}^{(t)}$ when employing the proposed gradient reconstruction algorithm with $R=3$ and $S_{\rm ratio}=8 \%$ at iteration $T=30$.
    Fig.~\ref{fig:Bernoulli} compares the \textit{empirical} cumulative distribution function (CDF) of the local gradient entries sampled from simulation with the CDF of the Bernoulli Gaussian-mixture distribution whose parameters are determined by {\bf Procedure \ref{alg:QEMGAMP}}.
    Samples $1$, $2$, and $3$ in Fig.~\ref{fig:Bernoulli} correspond to the gradient sub-vectors sampled when $(k,b) = (1,1)$, $(15,5)$, and $(30,10)$, respectively. 
	Our numerical example demonstrates that the empirical CDF of the local gradient sub-vectors are almost the same with the estimated CDF obtained based on our statistical model. 
	This result implies that the local gradient sub-vector is effectively modeled as an IID random vector with the Bernoulli Gaussian-mixture distribution.

	\subsection{Aggregate-and-Estimate Strategy}\label{Sec:BussGAMP}
    Although the estimate-and-aggregate strategy in Sec.~\ref{Sec:QGAMP} has a potential to minimize the MSE of local gradient estimates at the PS, the computational complexity of this strategy increases linearly with the number of wireless devices as the PS needs to perform the Q-EM-GAMP algorithm $KB$ times to reconstruct the global gradient vector in \eqref{eq:global_grad}.
        %The major reason is that the Q-EM-GAMP algorithm employed in the parallel recovery needs to be separately applied to each quantized sub-vector; thereby, this algorithm should be performed $KB$ times to reconstruct the global gradient vector in \eqref{eq:global_grad}.
    To mitigate this complexity requirement, we develop an aggregate-and-estimate strategy which promotes more flexibility on the complexity of the gradient reconstruction process. 
    %Unlike the parallel recovery, the joint recovery aims at reconstructing a {\em partial} aggregation of multiple local gradient sub-vectors, defined as

    The key idea of the aggregate-and-estimate strategy is to first aggregate a group of local gradient sub-vectors, then estimate the aggregated sub-vector. 
    In particular, we randomly divide $K$ devices into $G$ groups, then estimate an aggregated sub-vector for each group, defined as
    %to convert a global gradient reconstruction problem into $GB$ \textit{joint} recovery problems, each of which aims at reconstructing the aggregation of a group of the local gradient sub-vectors, defined as
    \begin{align}\label{eq:partial_grad}
        {\bf g}_{\mathcal{K}_g,b}^{(t)}  = \sum_{k\in\mathcal{K}_g} \rho_k^{(t)} {\bf g}_{k,b}^{(t)},~~\forall g\in\{1,\ldots,G\},
    \end{align}	  
    where $\mathcal{K}_g$ is the index set of the devices in the $g$-th group, and $\mathcal{K}_1,\ldots,\mathcal{K}_G$ are mutually exclusive subsets of $\mathcal{K}$ such that $\mathcal{K} = \bigcup_{g=1}^G \mathcal{K}_g$. 
    The underlying difficulty in this strategy is that aggregating the local gradient sub-vectors, $\{{\bf g}_{k,b}^{(t)}\}_{k\in\mathcal{K}_g}$, from the quantized sub-vectors, $\{{\bf q}_{k,b}^{(t)}\}_{k\in\mathcal{K}_g}$, is not straightforward due to nonlinearity of scalar quantization. 
        %To be more specific, a direct aggregation of {\em quantized} gradient sub-vectors does not lead to the quantization of the aggregation of the local gradient sub-vectors, e.g., $\sum_{k\in\mathcal{K}_g} \rho_k^{(t)} {\sf Q}\big({\bf x}_{k,b}^{(t)}\big)  \neq {\sf Q}\big( \sum_{k\in\mathcal{K}_g} \rho_k^{(t)} {\bf x}_{k,b}^{(t)}\big)$.
    To tackle this difficulty, we use the Bussgang theorem in \cite{Bussgang} which provides a theoretical basis to transform a quantization of a Gaussian signal into a linear signal with additive distortion.
    %Fortunately, each entry of ${\bf x}_{k,b}^{(t)}$ behaves like a Gaussian random variable with $\mathcal{N}(0,1)$ for large $N$, as discussed in {\bf Remark 1}.     
    Recall that each entry of ${\bf x}_{k,b}^{(t)}$ behaves like a Gaussian random variable with zero mean and unit variance for large $N$ by the central limit theorem, as discussed in Sec.~\ref{Sec:Comp_Block}.
    Meanwhile, different entries of ${\bf x}_{k,b}^{(t)}$ are uncorrelated as ${\bf A}$ is an IID random matrix.
    % because 
    % \begin{align}\label{eq:uncorr_x}
    %      \mathbb{E}\big[x_{k,b,m_1}^{(t)}x_{k,b,m_2}^{(t)} \big] =  \big(\alpha_{k,b}^{(t)}\big)^2\big({\bf g}_{k,b}^{(t)} \big)^{\sf T}\mathbb{E}\big[ {\bf a}_{{\rm row},m_1}^{\sf T} {\bf a}_{{\rm row},m_2}\big] {\bf g}_{k,b}^{(t)} \overset{(a)}{=} 0,
    % \end{align}
    % where the equality (a) holds because each element of ${\bf A}$ is independently drawn from $\mathcal{N}(0,1/M)$.
    %These facts allow us to adopt the Gaussian modeling of ${\bf x}_{k,b}^{(t)}  \sim \mathcal{N}({\bf 0}_{M},{\bf I}_{M})$ for all $k,b$.
        % We start by introducing the Bussgang theorem in \cite{Bussgang}, which is a key ingredient of the formulation of a joint gradient recovery problem.
        % In general, a direct aggregation of {\em quantized} vectors does not lead to the quantization of the aggregation of {\em unquantized} vectors due to the nonlinearity of the scalar quantizer ${\sf Q}(\cdot)$, e.g., $\sum_{k\in\mathcal{K}_g} \rho_k^{(t)} {\sf Q}\big({\bf x}_{k,b}^{(t)}\big)
        %     \neq {\sf Q}\big( \sum_{k\in\mathcal{K}_g} \rho_k^{(t)} {\bf x}_{k,b}^{(t)}\big)$.
        % To tackle this difficulty, we leverage the Bussgang theorem which provides a theoretical basis to transform a quantization of a Gaussian signal into a linear signal with additive distortion.
        % This theorem is particularly applicable in our setting because each entry of ${\bf x}_{k,b}^{(t)}$ behaves like a Gaussian random variable with $\mathcal{N}(0,1)$ for large $N$, as we already discussed in Sec.~\ref{Sec:Comp}. 
    Utilizing the modeling of ${\bf x}_{k,b}^{(t)}  \sim \mathcal{N}({\bf 0}_{M},{\bf I}_{M})$, we apply the Bussgang theorem to each quantized sub-vector, ${\bf q}_{k,b}^{(t)} = {\sf Q}\big({\bf x}_{k,b}^{(t)}\big)$, which yields the following proposition:
    
    \vspace{1mm}
    \begin{prop}\label{prop:Bussgang}
        Suppose that ${\sf Q}(x)$ is a scalar quantizer optimized for the distribution of $x\sim \mathcal{N}(0,1)$. If ${\bf x}_{k,b}^{(t)}  \sim \mathcal{N}({\bf 0}_{M},{\bf I}_{M})$, the following decomposition holds:
        \begin{align}\label{eq:Buss_block}
            {\sf Q}\big({\bf x}_{k,b}^{(t)}\big)  = \gamma_{Q}  {\bf x}_{k,b}^{(t)}  + {\bf d}_{k,b}^{(t)},
            %= \gamma_{Q}  \alpha_{k,b}^{(t)} {\bf A}{\bf g}_{k,b}^{(t)}  + {\bf d}_{k,b}^{(t)},
        \end{align}	
        where ${\bf d}_{k,b}^{(t)}$ is a quantization distortion uncorrelated with ${\bf x}_{k,b}^{(t)}$, and $\gamma_{Q}$ and $\psi_Q$ are quantizer-dependent constants defined as
        % \begin{align}\label{eq:gamma_psi_Q}
        %     \gamma_{Q} = \sum_{i=1}^{2^Q} \frac{q_i}{\sqrt{2\pi}} \left\{ \exp\left(-    \frac{\tau_{i-1}^2}{2}\right) - \exp\left(- \frac{\tau_{i}^2}{2}\right) \right\},~\text{and}~
        %     \psi_Q = \sum_{i=1}^{2^Q} q_i^2 \int_{\tau_{i-1}}^{\tau_i} \frac{1}{\sqrt{2\pi}} e^{-\frac{u^2}{2}} {\rm d}u.
        % \end{align}	
        \begin{align}\label{eq:gamma_Q}
            \gamma_{Q} = \sum_{i=1}^{2^Q} \frac{q_i}{\sqrt{2\pi}} \left\{ \exp\left(-    \frac{\tau_{i-1}^2}{2}\right) - \exp\left(- \frac{\tau_{i}^2}{2}\right) \right\},
        \end{align}	
        and     
        \begin{align}\label{eq:psi_Q}
            \psi_Q = \sum_{i=1}^{2^Q} q_i^2 \int_{\tau_{i-1}}^{\tau_i} \frac{1}{\sqrt{2\pi}} e^{-\frac{u^2}{2}} {\rm d}u,
            %\psi_Q = \sum_{i=1}^{2^Q} q_i^2  \left\{ \Phi(\tau_i) - \Phi(\tau_{i-1}) \right\}.
        \end{align}
        respectively.
        %with $\Phi(x) = \int_{-\infty}^{x} \frac{1}{\sqrt{2\pi}} e^{-\frac{u^2}{2}} {\rm d}u$.
        The distortion ${\bf d}_{k,b}^{(t)}$ has zero mean and the covariance of ${\bf  R}_{{\bf d}_{k,b}^{(t)} } = \big(\psi_Q - \gamma_{Q}^2\big)  {\bf I}_{M}$.
    \end{prop}
    \begin{IEEEproof}
        Let ${\bf q}_{k,b}^{(t)} = {\sf Q}\big({\bf x}_{k,b}^{(t)}\big)$. The Bussgang theorem in \cite{Bussgang} shows that if ${\bf x}_{k,b}^{(t)}  \sim \mathcal{N}({\bf 0}_{M},{\bf I}_{M})$, we have ${\bf q}_{k,b}^{(t)}  = \gamma_{Q}  {\bf x}_{k,b}^{(t)}  + {\bf d}_{k,b}^{(t)}$, where ${\bf d}_{k,b}^{(t)}$ is a quantization distortion uncorrelated with ${\bf x}_{k,b}^{(t)}$ and $\gamma_{Q}$ is a quantizer-dependent constant defined in \eqref{eq:gamma_Q} as derived in \cite{Jacobsson:17}.
        If the scalar quantizer ${\sf Q}(\cdot)$ is optimized for $\mathcal{N}(0,1)$, the distortion ${\bf d}_{k,b}^{(t)}$ has zero mean because both  ${\bf q}_{k,b}^{(t)}$ and  ${\bf x}_{k,b}^{(t)}$ have zero mean. 
        In addition, since ${\bf x}_{k,b}^{(t)}$ and ${\bf d}_{k,b}^{(t)}$ are uncorrelated, the covariance of ${\bf q}_{k,b}^{(t)}$ is obtained as ${\bf  R}_{{\bf q}_{k,b}^{(t)} } = \gamma_{Q}^2 {\bf I}_{M} + {\bf  R}_{{\bf d}_{k,b}^{(t)} }$,
        %${\bf  R}_{\hat{\bf q}_{k}^{(t)} } = \gamma_{Q}^2 {\bf I}_{M} + {\bf  R}_{{\bf d}_{k}^{(t)} }$.
        % \begin{align}\label{eq:cov_quant}
        %     {\bf  R}_{{\bf q}_{k,b}^{(t)} } = \gamma_{Q}^2 {\bf I}_{M} + {\bf  R}_{{\bf d}_{k,b}^{(t)} },
        % \end{align}	
        where ${\bf R}_{{\bf d}_{k,b}^{(t)}}$ is the covariance of ${\bf d}_{k,b}^{(t)}$.
        It is also easy to show that ${\bf  R}_{{\bf q}_{k,b}^{(t)} } = \psi_Q {\bf I}_{M}$ when ${\bf x}_{k,b}^{(t)}  \sim \mathcal{N}({\bf 0}_{M},{\bf I}_{M})$, where $\psi_Q$ is a quantizer-dependent constant defined in \eqref{eq:psi_Q}.
        Combining these results yields ${\bf  R}_{{\bf d}_{k,b}^{(t)} } = \big(\psi_Q - \gamma_{Q}^2\big)  {\bf I}_{M}$.
        % \begin{align}\label{eq:Cov_distortion}
        %     {\bf  R}_{{\bf d}_{k,b}^{(t)} } = \big(\psi_Q - \gamma_{Q}^2\big)  {\bf I}_{M}.
        % \end{align}
    \end{IEEEproof}
    \vspace{1mm}

    % we apply Bussgang's theorem to a quantized sub-vector, ${\bf q}_{k,b}^{(t)} = {\sf Q}\big({\bf x}_{k,b}^{(t)}\big)$, which yields \cite{Bussgang}
    % \begin{align}\label{eq:Buss_block}
    %     {\bf q}_{k,b}^{(t)} = \gamma_{Q}  {\bf x}_{k,b}^{(t)}  + {\bf d}_{k,b}^{(t)}
    %     = \gamma_{Q}  \alpha_{k,b}^{(t)} {\bf A}{\bf g}_{k,b}^{(t)}  + {\bf d}_{k,b}^{(t)},
    % \end{align}	
    % where ${\bf d}_{k,b}^{(t)} $ is a quantization distortion uncorrelated with ${\bf x}_{k,b}^{(t)}$, and $\gamma_{Q}$ is a quantizer-dependent constant defined as \cite{Jacobsson:17}
    % \begin{align}\label{eq:gamma_Q}
    %      \gamma_{Q} = \sum_{i=1}^{2^Q} \frac{q_i}{\sqrt{2\pi}} \left\{ \exp\left(- \frac{\tau_{i-1}^2}{2}\right) - \exp\left(- \frac{\tau_{i}^2}{2}\right) \right\}.
    % \end{align}	
    Utilizing the result in Proposition~\ref{prop:Bussgang}, we aggregate the quantized sub-vectors by assigning a special weight $\frac{\rho_k^{(t)}}{\gamma_Q\alpha_{k,b}^{(t)}}$ to ${\bf q}_{k,b}^{(t)}$ as follows:
    %formulate the problem of reconstructing ${\bf g}_{\mathcal{K}_g,b}^{(t)}$ in \eqref{eq:partial_grad} by computing the weight combination of the quantized sub-vectors using this weight:
    \begin{align}
        \tilde{\bf q}_{\mathcal{K}_g,b}^{(t)} = \sum_{k\in\mathcal{K}_g} \frac{\rho_k^{(t)}}{\gamma_Q\alpha_{k,b}^{(t)}} {\bf q}_{k,b}^{(t)} 
        &= \sum_{k\in\mathcal{K}_g} \frac{\rho_k^{(t)}}{\gamma_Q\alpha_{k,b}^{(t)}}  \big(\gamma_{Q}  \alpha_{k,b}^{(t)} {\bf A}{\bf g}_{k,b}^{(t)} + {\bf d}_{k,b}^{(t)}\big) 
        = {\bf A}{\bf g}_{\mathcal{K}_g,b}^{(t)} + \tilde{\bf d}_{\mathcal{K}_g,b}^{(t)},  \label{eq:joint_recovery}
    \end{align}	    
    where $\tilde{\bf d}_{\mathcal{K}_g,b}^{(t)} = \sum_{k\in\mathcal{K}_g} \frac{\rho_k^{(t)}}{\gamma_Q\alpha_{k,b}^{(t)}}  {\bf d}_{k,b}^{(t)}$ is an effective distortion uncorrelated with ${\bf g}_{\mathcal{K}_g,b}^{(t)}$.
    As can be seen in \eqref{eq:joint_recovery}, estimating the aggregated sub-vector, ${\bf g}_{\mathcal{K}_g,b}^{(t)}$, from the observation of $\tilde{\bf q}_{\mathcal{K}_g,b}^{(t)}$ is formulated a standard CS recovery problem with noise.
    Nevertheless, it is still difficult to find the exact MMSE solution of this problem as the distribution of $\tilde{\bf d}_{\mathcal{K}_g,b}^{(t)}$ is unknown in general. 
    %Unfortunately, the exact distribution of $\tilde{\bf d}_{\mathcal{K}_g,b}^{(t)}$ is unknown in general which prevents the PS from employing a well-designed CS recovery algorithm to solve the problem in \eqref{eq:joint_recovery}.
    To circumvent this difficulty, we assume that the correlation among distortions from different gradient sub-vectors is negligible which can be justified when local gradient sub-vectors from different devices are computed from different training samples.
    We then model the effective distortion as a Gaussian random vector with consistent mean and covariance. 
    Under this strategy, the effective distortion $\tilde{\bf d}_{\mathcal{K}_g,b}^{(t)}$ becomes an additive white Gaussian noise (AWGN) with the variance of
    %the distribution of $\tilde{\bf d}_{\mathcal{K}_g,b}^{(t)}$ is approximated by $\tilde{\bf d}_{\mathcal{K}_g,b}^{(t)}  \sim \mathcal{N}\big({\bf 0}_{M}, \xi_{\mathcal{K}_g,b}^{(t)} {\bf I}_{M}\big)$, where
    \begin{align}\label{eq:distortion_var}
        \nu_{g,b}^{(t)}  =\frac{\psi_Q - \gamma_{Q}^2}{\gamma_{Q}^2} \sum_{k\in\mathcal{K}_g} \bigg(\frac{\rho_k^{(t)}}{\alpha_{k,b}^{(t)}}\bigg)^2.
    \end{align}    
    %and $\kappa_Q = \frac{\psi_Q - \gamma_{Q}^2}{\gamma_{Q}^2}$.
    Based on the AWGN modeling of the effective distortion, we solve each CS recovery problem of \eqref{eq:joint_recovery} by employing the original EM-GAMP algorithm in \cite{EMGAMP}.
    Unlike the Q-EM-GAMP algorithm in the estimate-and-aggregate strategy, the original EM-GAMP algorithm is directly applied to an \textit{unquantized} linear observation of $\tilde{\bf q}_{\mathcal{K}_g,b}^{(t)}$ while assuming IID Bernoulli Gaussian-mixture prior for the aggregated sub-vector, ${\bf g}_{\mathcal{K}_g,b}^{(t)}$.
    %Note that the original EM-GAMP algorithm is the special case of the Q-EM-GAMP algorithm in {\bf Procedure~\ref{alg:QEMGAMP}}, obtained by setting $Q=\infty$ and $\nu_{d} = \nu_{g,b}^{(t)}$.
    %In addition,  the Bernoulli Gaussian-mixture distribution is adopted to model each aggregated sub-vector, ${\bf g}_{\mathcal{K}_g,b}^{(t)}$, instead of modeling each local gradient sub-vector, ${\bf g}_{k,b}^{(t)}$. 
    Another key difference is that in the problem of \eqref{eq:joint_recovery}, we have a \textit{noisy} observation due to the presence of the effective distortion $\tilde{\bf d}_{\mathcal{K}_g,b}^{(t)}$; thereby, the effect of the noise is taken into account when employing the EM-GAMP algorithm to solve \eqref{eq:joint_recovery}.
    %Details of the EM-GAMP algorithm utilized in our joint recovery strategy can be found in \cite{EMGAMP}.
    We denote the EM-GAMP algorithm utilized in the aggregate-and-estimate strategy by ${\sf EMGAMP}\big(\tilde{\bf q}_{\mathcal{K}_g,b}^{(t)},\nu_{g,b}^{(t)}, {\bf A}\big)$, where $\tilde{\bf q}_{\mathcal{K}_g,b}^{(t)}$ is a linear observation, $\nu_{g,b}^{(t)}$ is a AWGN variance, and ${\bf A}$ is a sensing matrix. 
    This EM-GAMP algorithm is also obtained by replacing Step 7 and Step 8 in {\bf Procedure~\ref{alg:QEMGAMP}} with $\hat{x}_m^{\rm post} = (\hat{p}_m\nu_{d}+q_m\nu_{p_m})/(\nu_{p_m}+\nu_{d})$ and $\nu_{x_m}^{\rm post} = (1/\nu_{p_m}+1/\nu_{d})^{-1}$, respectively, where $\nu_d$ is an input noise variance.
    % \begin{align}
    %     \hat{x}_m^{\rm post} &= (\hat{p}_m\nu_{d}+q_m\nu_{p_m})/(\nu_{p_m}+\nu_{d}),  \label{eq:Buss_x_post} \\
    %     \nu_{x_m}^{\rm post} &= (1/\nu_{p_m}+1/\nu_{d})^{-1}, \label{eq:Buss_nu_post} 
    % \end{align} 
    %respectively, where $\nu_d$ is an input noise variance.
    Finally, we reconstruct the global gradient vector by aggregating the estimates of the aggregated sub-vectors.
	Let $\hat{\bf g}_{\mathcal{K}_g,b}^{(t)}$ be the estimate of ${\bf g}_{\mathcal{K}_g,b}^{(t)}$ computed by applying the EM-GAMP algorithm to $\tilde{\bf q}_{\mathcal{K}_g,b}^{(t)}$. 
	Then the global gradient vector in \eqref{eq:global_grad} is reconstructed as $\hat{\bf g}_{\mathcal{K}}^{(t)} = \sum_{g=1}^G \hat{\bf g}_{\mathcal{K}_g}^{(t)}$,
% 	\begin{align}
%         \hat{\bf g}_{\mathcal{K}}^{(t)} = \sum_{g=1}^G \hat{\bf g}_{\mathcal{K}_g}^{(t)},
%     \end{align}	
    where $\hat{\bf g}_{\mathcal{K}_g}^{(t)} = {\sf Concatenate}\big(\{\hat{\bf g}_{\mathcal{K}_g,b}^{(t)}\}_{b=1}^B\big)$.
	The overall gradient reconstruction process of the aggregate-and-estimate strategy is summarized in Steps 16--20 of {\bf Procedure~\ref{alg:FedQCS}}.
	%In Step 18, ${\sf EMGAMP}({\bf q},\nu_d)$ represents the original EM-GAMP algorithm in \cite{EMGAMP} applied to a linear observation ${\bf q}$ with an input noise variance $\nu_d$. 
    
    % The overall gradient reconstruction process via the joint recovery strategy is summarized in Steps 16--20 of {\bf Procedure~\ref{alg:FedQCS}}.
    % In Step 18, ${\sf EMGAMP}({\bf q},\nu_d)$ represents the original EM-GAMP algorithm in \cite{EMGAMP} applied to a linear observation ${\bf q}$ with an input noise variance $\nu_d$. This algorithm is also the special case of the Q-EM-GAMP algorithm in {\bf Procedure~\ref{alg:QEMGAMP}}, obtained by replacing Step 7 and Step 8 with 
    % \begin{align}
    %     \hat{x}_m^{\rm post} &= (\hat{p}_m\nu_{d}+q_m\nu_{p_m})/(\nu_{p_m}+\nu_{d}),  \label{eq:Buss_x_post} \\
    %     \nu_{x_m}^{\rm post} &= (1/\nu_{p_m}+1/\nu_{d})^{-1}, \label{eq:Buss_nu_post} 
    % \end{align} 
    % respectively.
    % In Step 19, the aggregated gradient vector for each group, $\hat{\bf g}_{\mathcal{K}_g}^{(t)}$, is obtained by concatenating the associated gradient sub-vectors, $\{\hat{\bf g}_{\mathcal{K}_g,b}^{(t)}\}_{b=1}^B$, $\forall g$.
    % In Step 20, the global gradient vector in \eqref{eq:global_grad} is reconstructed by aggregating the aggregated gradient vectors for all groups, $\{\hat{\bf g}_{\mathcal{K}_g}^{(t)}\}_{g=1}^G$.

    A key benefit of the aggregate-and-estimate strategy is that it requires a lower complexity than the estimate-and-aggregate strategy when $G<K$ because the aggregate-and-estimate strategy solves only $GB$ CS recovery problems to reconstruct the global gradient vector.
    %Moreover, this strategy allows the PS to control the complexity of the gradient recovery process by adjusting the number of groups, $G$.  
    % In the joint strategy, the PS needs to perform a recovery algorithm only $GB$ times to reconstruct the global gradient vector; thereby, this strategy requires a less complexity than the parallel recovery strategy in Sec.~\ref{Sec:QGAMP}.
    Another key benefit is that the aggregate-and-estimate strategy allows the PS to control the complexity of the gradient reconstruction process by adjusting the number of sub-vectors per group. 
    It is worth mentioning that the complexity reduction achieved by this strategy comes at the cost of reconstruction accuracy because the number of non-zero values in ${\bf g}_{\mathcal{K}_g,b}^{(t)}$ is higher than that in ${\bf g}_{k,b}^{(t)}$, which degrades the accuracy of the CS recovery process. 
    In particular, the degradation in the reconstruction accuracy becomes severe as the number of sub-vectors per group increases because the larger the number of the sub-vectors per group, the larger the number of non-zero values in ${\bf g}_{\mathcal{K}_g,b}^{(t)}$.
    Therefore, when employing the aggregate-and-estimate strategy, there is a trade-off between the accuracy and the complexity of the gradient reconstruction.

    \section{Performance Analysis}\label{Sec:Analysis}
    In this section, we analyze the gradient reconstruction error as well as the convergence rate of FedQCS. 
    We first characterize an upper bound for the reconstruction error achieved by FedQCS with the aggregate-and-estimate strategy in Sec.~\ref{Sec:BussGAMP}.
    We then characterize the convergence rate of FedQCS under the consideration of the reconstruction error bound.

    \subsection{Reconstruction Error Analysis}\label{Sec:ErrorAnalysis}
    In this analysis, we characterize an upper bound of the gradient reconstruction error in FedQCS.
    We particularly aim at analyzing the performance of the aggregate-and-estimate strategy with $G=1$ which provides the worst-case performance as discussed in Sec.~\ref{Sec:BussGAMP}.
    %Note that the joint recovery with $G=1$ provides the wort reconstruction error among all possible reconstruction strategies of FedQCS, as discussed in Sec.~\ref{Sec:BussGAMP}.    %As we already discussed in Sec.~\ref{Sec:Recovery}, the reconstruction error of the Buss-EM-GAMP algorithm in Sec.~\ref{Sec:BussGAMP} increases as $G$ decreases, while it performs worse than the Q-EM-GAMP algorithm. 
        %Motivated by this observation, we aim at characterizing the MSE bound of the Buss-EM-GAMP algorithm with $G=1$, in order to guarantee the convergence of FedQCS even in the worst case. 
    %Furthermore, since the reconstruction error analysis is mathematically intractable in general federated learning scenarios, we resort to some useful assumptions: 
    We also make some useful assumptions for mathematical tractability of the reconstruction error analysis, even if these assumptions are not necessary for employing the proposed gradient reconstruction strategies.
    %because the reconstruction error analysis for general federated learning scenarios is too complicated. 
    The assumptions made in our analysis are described below. 

    \vspace{1mm}
    {\em Assumption 1:} 
    Every local gradient sub-vector, ${\bf g}_{k,b}^{(t)}$, follows a Bernoulli Gaussian-mixture distribution in \eqref{eq:BernoulliGM} which is already known at the PS. 

    {\em Assumption 2:} 
    Every quantization distortion, ${\bf d}_{k,b}^{(t)}$, follows a Gaussian distribution. 
    \vspace{1mm}
 
    %Although the above assumptions are not necessary for employing the gradient reconstruction strategies proposed in our work, these assumptions are useful to provide the mathematical tractability of the reconstruction error analysis.
    Under Assumptions 1 and 2, we characterize an upper bound of the MSE of the global gradient vector as given in the following theorem:

    \vspace{1mm}
    \begin{thm}\label{Thm1}
        Suppose that Assumptions 1 and 2 hold. 
        In the asymptotic regime of $N \rightarrow \infty$ and $N/M \rightarrow R$ for a fixed ratio $R\geq 1$, the global gradient vector reconstructed by the aggregate-and-estimate strategy with $G=1$ satisfies the following MSE bound:
        \begin{align}\label{eq:Thm1}
            \mathbb{E}\big[\| {\bf g}_{\mathcal{K}}^{(t)} - \hat{\bf g}_{\mathcal{K}}^{(t)}\|^2\big]
            \leq  N  \sum_{b=1}^B \tilde{\nu}_{\mathcal{K},b}^{(t)} \left( 1 - \frac{\tilde{\nu}_{\mathcal{K},b}^{(t)}}
            { R \tilde{\nu}_{\mathcal{K},b}^{(t)}  + \kappa_Q \big(\tilde{\nu}_{\mathcal{K},b}^{(t)} + \tilde{\mu}_{{\rm sq},\mathcal{K},b}^{(t)}  \big) } \right),
        \end{align}
        where  $\tilde{\mu}_{{\rm sq},\mathcal{K},b}^{(t)} = \sum_{k=1}^K \big( \rho_k^{(t)} \mu_{{\bf g}_{k,b}}^{(t)}\big)^2$ and $\tilde{\nu}_{\mathcal{K},b}^{(t)} =  \sum_{k=1}^K \big(\rho_k^{(t)}\big)^2\nu_{{\bf g}_{k,b}}^{(t)}$, provided that ${\bf g}_{k,b}^{(t)}$ is an IID random vector with mean $\mu_{{\bf g}_{k,b}}^{(t)}$ and variance $\nu_{{\bf g}_{k,b}}^{(t)}$. 
    \end{thm}
    \begin{IEEEproof}
        See Appendix~\ref{Apdx:Thm1}.
    \end{IEEEproof}
    \vspace{1mm}

    Theorem~1 shows how the reconstruction error in FedQCS depends on the dimension reduction ratio $R$, the quantization function (captured by $\kappa_Q$), and the distribution of the gradient vector (captured by $\tilde{\mu}_{{\rm sq},\mathcal{K},b}^{(t)}$ and $\tilde{\nu}_{\mathcal{K},b}^{(t)}$).
    %In particular, the impact of the quantization function reduces as $Q$ increases because $\kappa_Q \rightarrow 0$ as $Q \rightarrow \infty$.
    Since $\kappa_Q \rightarrow 0$ as $Q \rightarrow \infty$, Theorem~1 also demonstrates that the reconstruction error in FedQCS vanishes as $Q \rightarrow \infty$ and $R \rightarrow 1$. 
    This result implies that perfect reconstruction of the global gradient vector is feasible at the PS when employing FedQCS with $Q =\infty$ and $R=1$, which also coincides with our intuition.

    \subsection{Convergence Rate Analysis}
    In this analysis, we characterize the convergence rate of FedQCS operating with the SGD algorithm. 
    We particularly make the following assumptions not only to provide mathematical tractability for the convergence rate analysis, but also to connect this analysis with the reconstruction error analysis in Sec.~\ref{Sec:ErrorAnalysis}.
    
    \vspace{1mm}
    {\em Assumption 3:} 
    The loss function $F({\bf w})$ is $\beta$-smooth and is lower bounded by some constant $F({\bf w}^{\star})$, i.e.,  $F({\bf w})\geq F({\bf w}^{\star})$, $\forall {\bf w}\in \mathbb{R}^{\bar{N}}$. 

    {\em Assumption 4:} 
    For a given parameter vector ${\bf w}_t$, the global gradient vector in \eqref{eq:global_grad} is unbiased and has bounded variance, i.e., $\mathbb{E} \big[{\bf g}_{\mathcal{K}}^{(t)} \big| {\bf w}_t \big] = \nabla F({\bf w}_t)$ and
    $\mathbb{E} \big[\|{\bf g}_{\mathcal{K}}^{(t)} - \nabla F({\bf w}_t)\|^2 \big| {\bf w}_t \big] \leq \sigma^2$, for all $t\in\{1,\ldots,T\}$. 
    
    {\em Assumption 5:} 
    The squared reconstruction error is upper bounded by the squared norm of the true gradient vector with some scaling factor $\epsilon <1$, i.e.,  
    $\| \hat{\bf g}_{\mathcal{K}}^{(t)} - {\bf g}_{\mathcal{K}}^{(t)} \|^2 \leq \epsilon \|\nabla F({\bf w}_t) \|^2$, for all $t\in\{1,\ldots,T\}$. 
    \vspace{1mm}

    We would like to make some important comments on the above assumptions. 
    Assumption 3 is standard for analyzing the convergence properties of a family of gradient descent algorithms (e.g., \cite{SignSGD,Lee:21}).
    Assumption 4 is useful to capture the impact of the mini-batch size as well as the block sparsification level $S$. 
    A similar assumption is also considered in the literature (e.g., \cite{SignSGD}). 
    More precisely, the variance bound $\sigma^2$ can be made smaller by increasing both the mini-batch size and the sparsification level. 
    Assumption 5 is particularly relevant to FedQCS because the recovery strategies in FedQCS aim at reducing the squared estimation error of the global gradient vector. 
    From Theorem 1, we have already shown that the expectation of the squared reconstruction error vanishes as $Q \rightarrow \infty$ and $R \rightarrow 1$, implying that $\epsilon \rightarrow 0$ as $Q \rightarrow \infty$ and $R \rightarrow 1$. 
    In Sec.~\ref{Sec:Simul}, we will also demonstrate that the scaling factor $\epsilon$ is extremely small under practical scenarios (i.e., $\epsilon \ll 1$).

    Under Assumptions $3\sim 5$, we characterize the convergence rate of FedQCS with a fixed learning rate $\eta_t = \frac{(1-\sqrt{\epsilon})} {2\beta (1+\epsilon) \sqrt{T}}$, as given in the following theorem:

    \vspace{1mm}
    \begin{thm}\label{Thm2}
        Under Assumptions $3\sim5$, FedQCS with a fixed learning rate $\eta_t = \frac{(1-\sqrt{\epsilon})} {2\beta (1+\epsilon) \sqrt{T}}$ satisfies the following bound:
        \begin{align}\label{eq:Thm2}
            \mathbb{E} \left[\frac{1}{T} \sum_{t=1}^T  \| \nabla F({\bf w}_t) \|^2\right]
            \leq \frac{1}{\sqrt{T}} \left[  \frac{4\beta (1+ \epsilon) }{ (1-\sqrt{\epsilon})^2}\big\{ F({\bf w}_1) - F({\bf w}^*)\big\} + \frac{\sigma^2}{1+\epsilon} \right].
        \end{align}
    \end{thm}
    \begin{IEEEproof}
        See Appendix~\ref{Apdx:Thm2}.
    \end{IEEEproof}
    \vspace{1mm}

    Theorem~2 demonstrates that FedQCS converges to a stationary point of the loss function if the initial loss, $F({\bf w}_1) - F({\bf w}^{\star})$, the variance of the global gradient vector (captured by $\sigma^2$), and the reconstruction error (captured by $\epsilon$) are finite. 
    It is also shown that the convergence rate of FedQCS has the order of $\mathcal{O}\big(\frac{1}{\sqrt{T}}\big)$ which is the same as that of the original SGD algorithm. 
    The scaling factor of the convergence rate decreases as both $\epsilon$ and $\sigma^2$ reduces; thereby, the convergence rate of FedQCS improves as $Q \rightarrow \infty$ and $R \rightarrow 1$ provided that both the mini-batch size and the sparsification level are sufficiently large.

    \section{Simulation Results}\label{Sec:Simul}
    In this section, we demonstrate the superiority of FedQCS over the existing federated learning frameworks, using simulations.
 	%\subsection{Federated Learning Scenario}\label{Sec:FL-Scenario}
% 	In this section, we describe the simulation settings for evaluating the performance of algorithms described in above. 
	In these simulations, we consider an image classification task using the publicly accessible MNIST dataset, where each data sample is a $28 \times 28$ grayscale image representing a handwritten digit from $0$ to $9$ \cite{MNIST}. 
	The MNIST dataset consists of the $60,000$ training data samples and the $10,000$ test data samples. 
	We set the number of wireless devices as $K=30$ and consider {\em non-IID} distribution of the training data samples over the devices. 
	%assume that each device has the training data samples associated with only one digit; this corresponds to {\em non}-IID setting.
	In particular, we construct the local training data set of device $k$, $\mathcal{D}_k$, by randomly selecting $1,000$ training data samples labeled with $d_k = \big\lfloor \frac{k-1}{K/10} \big\rfloor$ in the MNIST dataset. 
	%which are used to test the classification accuracy of the trained global model. 
	A global model on the PS is assumed to be a neural network that consists of $784$ input nodes, a single hidden layer with $20$ hidden nodes, and $10$ output nodes. 
	The activation functions of the hidden layer and the output layer are set as the rectified linear unit and the softmax function, respectively. 
	The total number of the weights in the global model is $\bar{N} = 15,910$. 
	To train the global model, we adopt the Adam optimizer in \cite{ADAM} with a learning rate $0.003$ and cross-entropy loss function.
	%We also assume that the Adam optimizer in \cite{ADAM} with a learning rate $0.003$ is employed at the PS to train the global model, where the loss function is set as cross entropy.
	%This corresponds to {\em non}-IID setting because each device has the information of only one digit. 
	We also consider {\em stochastic} gradient descent setting by assuming that $|\mathcal{D}_k^{(t)}| = 1$, for all $k\in\mathcal{K}$ and $t \in \{1,\ldots,T\}$.
	
    For performance evaluation, we mainly consider two performance metrics: (i) classification accuracy and (ii) normalized MSE (NMSE), defined as ${\big\|{\bf g}^{(t)}_{\mathcal{K}}-\hat{\bf g}^{(t)}_{\mathcal{K}}\big\|^2}/{\big\|{\bf g}^{(t)}_{\mathcal{K}} \big\|^2}$. 
    % To demonstrate the superiority of the proposed algorithms (Q-EM-GAMP and Buss-EM-GAMP), we conduct the experiments on the federated learning system with $K = 30$ devices. In these experiments, we follow the block sparsification described in Sec.~\ref{Sec:Comp_Block} with $M = \lfloor N/R \rfloor$ and $S = \lfloor S_{\rm ratio}/R \rfloor$, where $S_{\rm ratio}$ is the ratio of the sparsification level in the block sparsification process. 
    Federated learning frameworks considered for performance comparison are described below.
	\begin{itemize}
        \item {\em FedQCS-EA and FedQCS-AE:} 
        %The proposed algorithms are Q-EM-GAMP and Buss-EM-GAMP, where the details of each algorithm are described in {\bf Procedure~\ref{alg:QEMGAMP}} and {\bf Procedure~\ref{alg:BussEMGAMP}}, respectively. 
        %It is noted that both algorithms leverage the prior knowledge about the distribution of each entry in gradient vector, namely the Bernoulli Gaussian-mixture model $\mathcal{BG}(g;{\bm \theta})$ in \eqref{eq:BernoulliGM}. 
        %From this reason, both algorithms takes the same initialization for the model parameters ${\bm \theta} = (\lambda_0,\{\lambda_l,\mu_l,\phi_l\}_{l=1}^L)$. %In the implementation of both algorithms, 
        FedQCS-EA and FedQCS-AE are the proposed FedQCS using the estimate-and-aggregate strategy in Sec.~\ref{Sec:QGAMP} and the aggregate-and-estimate strategy in Sec.~\ref{Sec:BussGAMP}, respectively.
        The Q-EM-GAMP algorithm and the EM-GAMP algorithm adopted in these strategies are initialized as follows:
        We randomly set an initial estimate as $\hat{g}_n \sim \mathcal{N}(0, \frac{M}{N\alpha^2})$, $\forall n\in\{1,\ldots,N\}$, considering the fact that $\alpha^2 = M/{\|{\bf g}\|^2}$.
        We then initialize the parameters ${\bm \theta}$ of the Bernoulli Gaussian-mixture model as $L = 3$, $\lambda_0 = 0.9$, $\lambda_l = \frac{1-\lambda_0}{L}$, $\mu_l = \hat{g}_{\rm min} + \frac{2l-1}{2L}(\hat{g}_{\rm max} - \hat{g}_{\rm min})$, $\phi_l = \frac{1}{12}\big(\frac{\hat{g}_{\rm max} - \hat{g}_{\rm min}}{L}\big)^2$, $\forall l \in \{1,\ldots,L\}$, where $\hat{g}_{\rm max} = \underset{n}{\max}~g_n$ and $\hat{g}_{\rm min}= \underset{n}{\min}~g_n$. 
        We also set $\tau_{\rm GAMP} = 10^{-5}$ and $I_{\rm GAMP} = 50$ for the stopping criterion.

        \item {\em QCS-QIHT:} 
        QCS-QIHT is a simple modification of the QCS-based federated learning framework introduced in \cite{QCS1,QCS2}.
        In this modification, we employ the estimate-and-aggregate strategy based on the quantized iterative hard thresholding (QIHT) algorithm in \cite{QIHT}, instead of the Q-EM-GAMP algorithm, while utilizing the proposed BQCS compression in Sec.~\ref{Sec:Comp_Block}.
        It is worth mentioning that unlike the Q-EM-GAMP algorithm, the QIHT algorithm requires the knowledge of the sparsity level $S$.
        %thereby, this algorithm cannot be extended to the estimate-and-aggregate strategy. 
        In this algorithm, we scale the reconstructed gradient vector, $\hat{\bf g}_{k,b}^{(t)}$,  using the scaling factor $\alpha_{k,b}^{(t)}$ in order to make the norm of $\hat{\bf g}_{k,b}^{(t)}$ consistent with that of the true gradient vector ${\bf g}_{k,b}^{(t)}$.

        \item {\em QCS-Dither:} 
        QCS-Dither is the existing QCS-based federated learning framework introduced in \cite{DitherQCS}.
        % The algorithm in \cite{DitherQCS}, namely unbiased QCS, is based on the compressed sensing and dither quantization theories \cite{DitherTheory}. The dither quantization theory demonstrates that the quantized signal can be decomposed into the sum of the quantizer input signal and independent quantization distortion if two necessary conditions are met, where its two conditions are (i) the use of uniform quantizer and (ii) the sharing of random dither signal with devices and PS. Furthermore, there are other notable points in this algorithm as below: 
        This framework employs the gradient compression based on dithered uniform quantization with dimension reduction, while utilizing a simple linear estimator to reconstruct the global gradient vector. 
        Unlike FedQCS and QCS-QIHT, the sensing matrix ${\bf A}$ is determined as the product of the Hadamard and random Rademacher diagonal matrix, as proposed in \cite{DitherQCS}.
        %for the compression of gradient vector, whereas other QCS algorithms as well as the proposed algorithms represented in this paper make use of the Gaussian matrix $\bf A \sim \mathcal{N}(0,I)$. Second, the reconstruction of gradient vector is restricted to using linear estimator, specifically the transposed compression matrix ${\bf A}^{\rm T}$.
        %In this algorithm, we scale the reconstructed gradient vector, $\hat{\bf g}_{k,b}^{(t)}$,  using the scaling factor $\alpha_{k,b}^{(t)}$ in order to make the norm of $\hat{\bf g}_{k,b}^{(t)}$ consistent with that of the true gradient vector ${\bf g}_{k,b}^{(t)}$. 
        %
        %Last, the  sparsification  process is not adopted, i.e, $S_{\rm ratio} = 100\,\%$.
        %In this algorithm, we scale the estimated gradient vector using the scaling factor $\alpha$ for the same reason as shown in the description of $\bf QIHT$ above.

        \item {\em SignSGD:} 
        SignSGD is the existing federated learning framework introduced in \cite{SignSGD}. 
        In SignSGD, each device transmits the sign of each entry of the local gradient vector,  then the PS aggregates the received signs of the local gradient entries by a majority vote. 
        For this reason, the communication overhead of SignSGD is $\bar{N}$ bits per device.
    	%\textcolor[rgb]{1,0,0}{In the one-bit quantization algorithm (i.e., SignSGD), we set the optimizer as SGD with a learning rate $0.0014$, where we get the highest classification accuracy using this learning rate in our experiments.} 
        %In the implementation of SignSGD algorithm, we assume that ${\rm sign}(0)$ takes a value $1$ with probability $1/2$, and $-1$ with probability $1/2$. It is clearly demonstrates that the one-bit quantization algorithms, including SignSGD, have a communication overhead per each device as $\bar{N}$ bits. This overhead can be achieved by QCS algorithms that adopt a ratio of $R/Q=1$. The numerical results comparing QCS algorithms with SignSGD are described in below.
    \end{itemize}
    Except for SignSGD, we adopt the block sparsification process with $B=10$
    %for MNIST, $B=30$ for CIFAR-$10$, 
    and $S = \lfloor S_{\rm ratio}N \rfloor$, where $S_{\rm ratio}$ is the ratio of the number of non-zero gradient entries.
    For FedQCS and QCS-QIHT, we adopt the Lloyd-Max scalar quantizer optimized for $\mathcal{N}(0,1)$. 
    %In the following simulations, we try to set the best value of $S_{\rm ratio}$ that maximizes the classification accuracy. 

	\begin{figure}
		\centering 
		\subfigure[Classification accuracy]
		{\epsfig{file=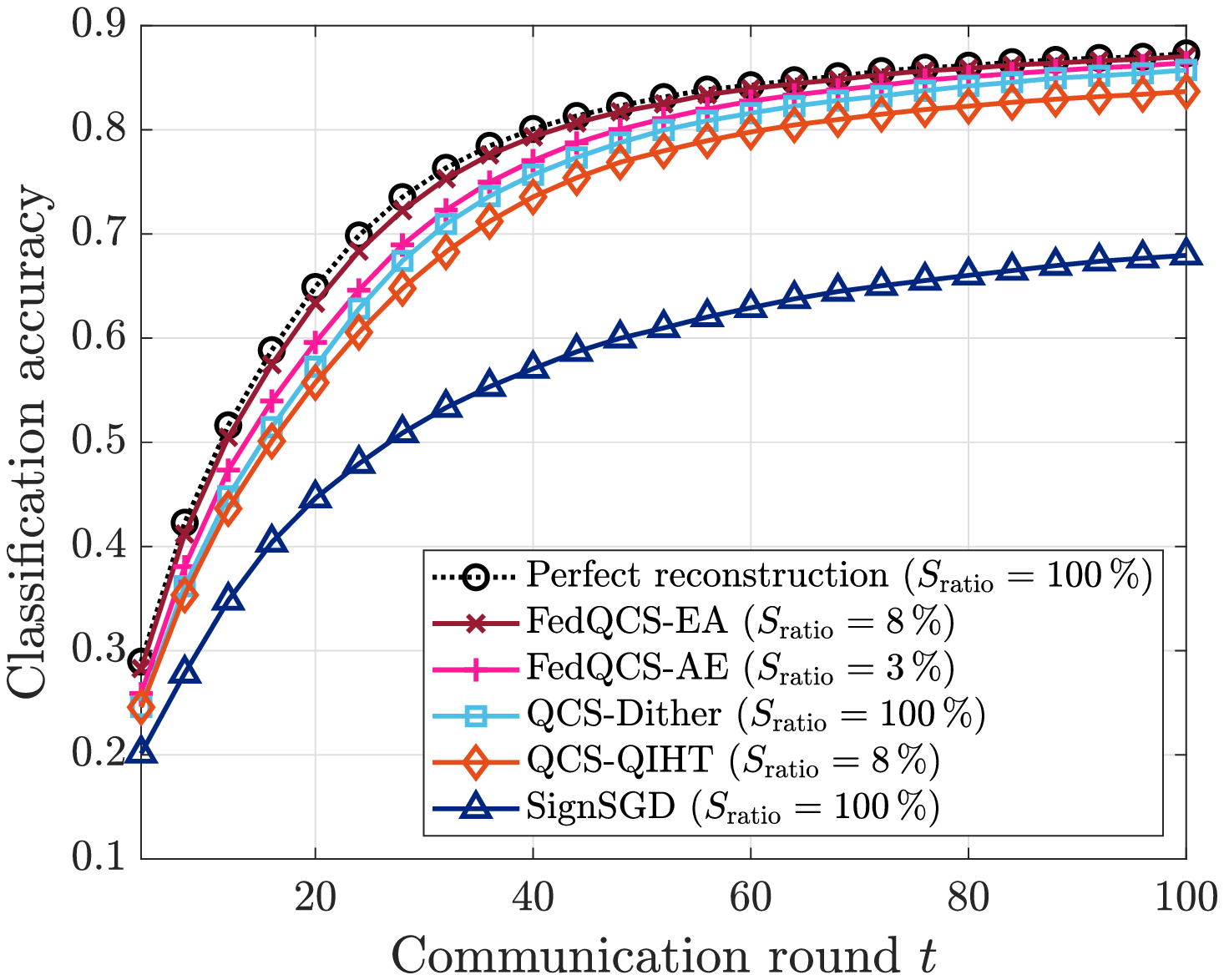, width=7.7cm}}
		%\qquad\qquad\qquad
		\subfigure[Normalized MSE]
		{\epsfig{file=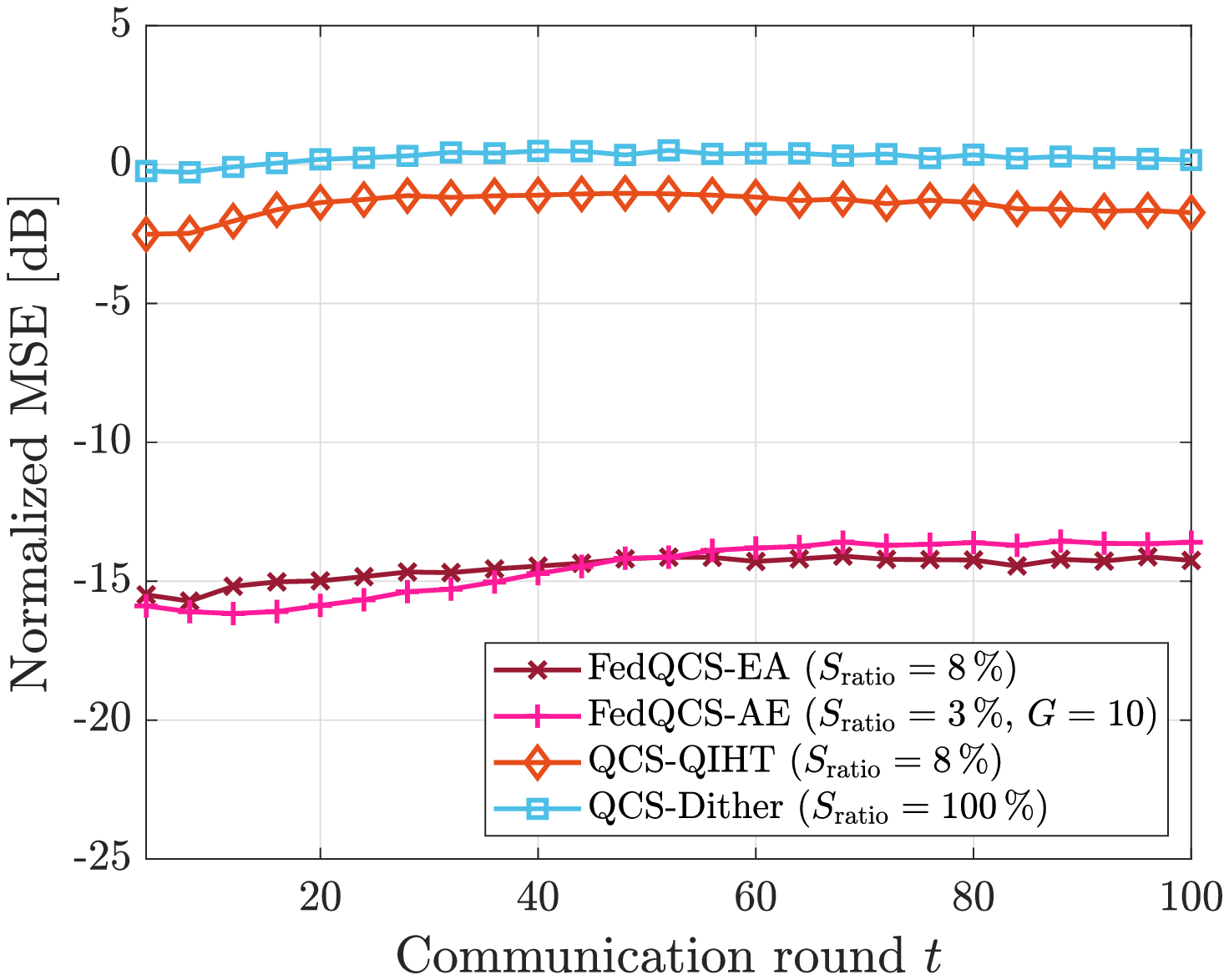, width=7.7cm}} \vspace{-2mm}
		\caption{Performance comparison of different federated learning frameworks with one bit overhead per gradient entry.} \vspace{-5mm}
		\label{fig:MNIST}
	\end{figure}

	%\subsection{Performance Comparison}\label{Sec:Performance}
	In Fig.~\ref{fig:MNIST}, we compare the classification accuracy and the NMSE of different federated learning frameworks with one bit overhead per gradient entry (i.e., $\bar{N}$ bits per device). In this simulation, we set $(R,Q)=(3,3)$ for all the QCS-based frameworks.
	%\textcolor[rgb]{0,0,1}{In this simulation, we use MNIST dataset and let $(R,Q)=(3,3)$ for all the QCS-based frameworks.}
	%We set the best values of $S_{\rm ratio}$ that maximizes the classification accuracy, which is specified in Fig.~\ref{fig:MNIST}.
	As a performance benchmark, we also plot the optimal performance achieved when the global gradient vector is perfectly reconstructed at the PS, requiring the communication overhead of $32$ bits per gradient entry when employing floating-point representation.
	Fig.~\ref{fig:MNIST}(a) shows that FedQCS-EA achieves the almost same accuracy as perfect reconstruction while requiring $32$ times less communication overhead.
	This result implies that the proposed federated learning framework enables not only significant reduction in the communication overhead, but also {\em almost lossless} reconstruction of the global gradient vector at the PS. %while achieving significant reduction in the communication overhead.
	It is also shown that both FedQCS-EA and FedQCS-AE achieve a higher classification accuracy (see Fig.~\ref{fig:MNIST}(a)) as well as a lower NMSE (see Fig.~\ref{fig:MNIST}(b)) compared to the existing frameworks requiring the same overhead.
	The performance gain of FedQCS over both QCS-QIHT and QCS-Dither demonstrates the superiority of the proposed recovery strategies over the QIHT algorithm in \cite{QCS1,QCS2} or a simple linear estimator in \cite{DitherQCS}. %comes from accurate reconstruction of the global gradient vector (see Fig.~\ref{fig:MNIST}(b))  	This result
	%Meanwhile, additional gain is attained over QCS-Dither because FedQCS utilizes the optimal scalar quantizer that minimizes quantization error, while QCS-Dither utilizes the dithered quantization with a uniform quantizer which is suboptimal.
	Meanwhile, the performance gain of FedQCS over SignSGD demonstrates that the proposed BQCS compression is more effective than one-bit gradient compression by exploiting the sparse property of the local gradients. 

\begin{figure}
        \centering
            \begin{minipage}{0.48\columnwidth}
            \centering
            {\setlength{\fboxrule}{0pt}
            \fbox{{\epsfig{file=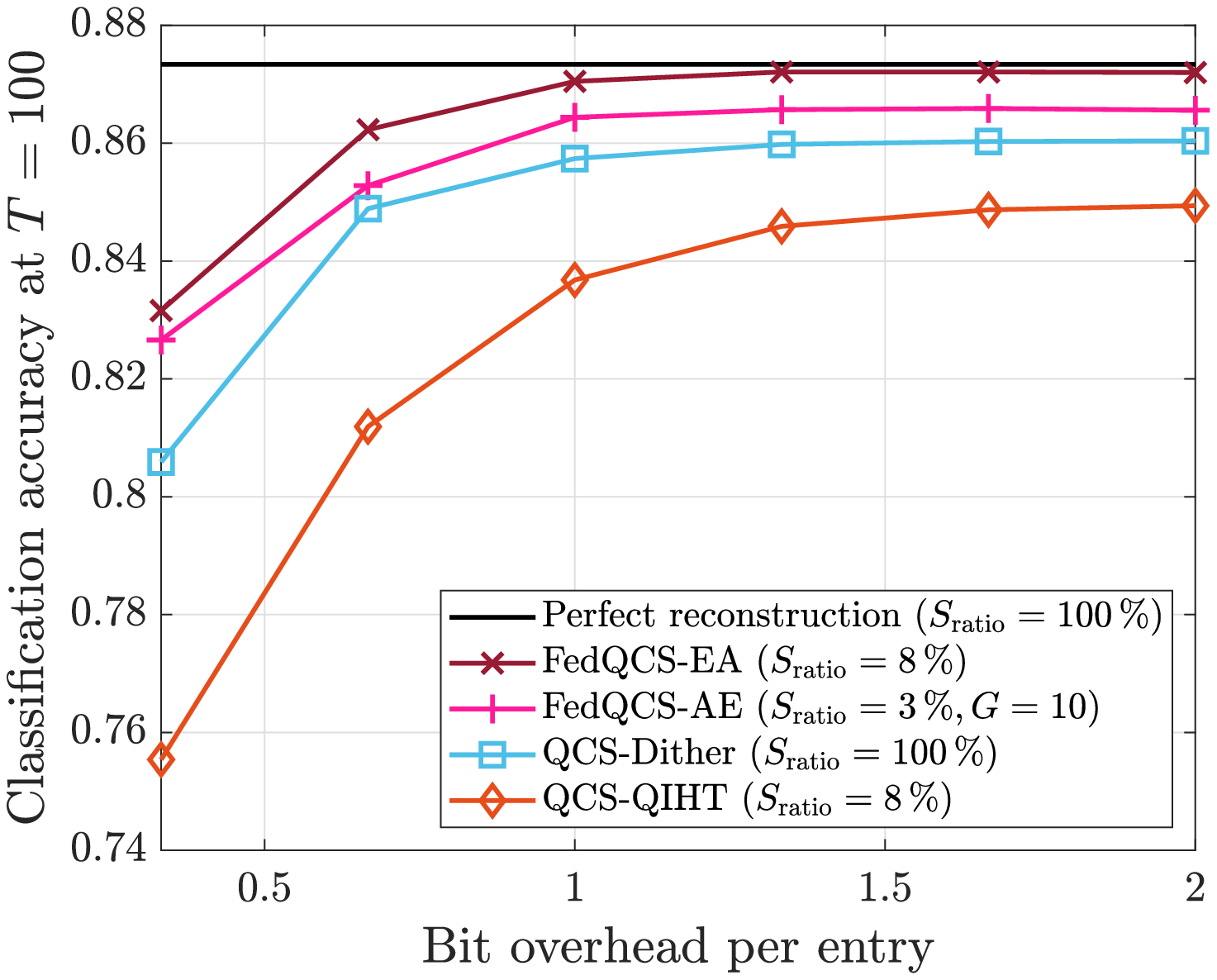, width=7.7cm}}}
            \vspace{-12mm}
            \captionof{figure}{The classification accuracy vs. communication overhead for different federated learning frameworks.}
            \vspace{-5mm}
            \label{fig:ACC_Bit}}
        \end{minipage}
         \hspace{1mm}
        \begin{minipage}{0.48\columnwidth}
            \centering
            \small
            \captionof{table}{Complexity order required by different QCS-based federated learning frameworks}
            {\setlength{\fboxrule}{0pt}
            \fbox{ \begin{tabular}{|c|c|}
            \hline Algorithm & Complexity Order \\ \hline \hline QCS-Dither & $\mathcal{O}(BMN)$ \\ \hline
            QCS-QIHT & $\mathcal{O}(KBMNI_{\rm QIHT})$ \\ \hline
            FedQCS-EA  & $\mathcal{O}(KBMNI_{\rm GAMP})$ \\ \hline
            FedQCS-AE  & $\mathcal{O}(GBMNI_{\rm GAMP})$ \\           \hline\end{tabular}}}
            \label{table:Complexity}
        \end{minipage}
    \end{figure}

	In Fig.~\ref{fig:ACC_Bit}, we compare the classification accuracy of different QCS-based federated learning frameworks with various communication overheads. In this simulation, we set $R=3$ and increase the value of $Q$ from $1$ to $6$, which corresponds to $1/3$-bit overhead to $2$-bit overhead per gradient entry. Fig.~\ref{fig:ACC_Bit} shows that FedQCS achieves the highest classification accuracy regardless of the communication overhead.
	Meanwhile, the performance gain of FedQCS over the existing frameworks increases as the communication overhead reduces. This result demonstrates the robustness of FedQCS against the increase in the compression ratio. 
	%This implies that FedQCS is robust the parallel recovery achieves almost the same classification accuracy as perfect reconstruction even with one bit overhead per gradient entry.
	%This implies that the use of FedQCS reduces communication overhead of federated learning   
	%while outperforming other federated learning frameworks regardless of the communication overhead. 
	%In particular, the classification accuracy of FedQCS with the parallel recovery with only $2/3$-bit outperforms that of both QCS-Dither and QCS-QIHT with $2$-bit. 
	Another interesting observation is that the classification accuracy of all the frameworks saturates as the communication overhead increases. 
	%with some value as the bit overhead increases. 
	This phenomenon implies that if the number of quantization bits becomes sufficiently large (e.g., $Q=4$), there is no significant reduction in the quantization error. %does not reduces, as also observed in \cite{FLBeam}.  
	%Therefore, under the same communication overhead determined by the ratio of $Q/R$, a judicious optimization of $R$ and $Q$ is still necessary in order to maximize the performance of the QCS-based federated learning. 
	%the quantization error tends to be much smaller than the error of compressed sensing when the bit overhead is larger than $1$-bit. In other words, 
	%even if the value of $Q$ is increased to $4$ or more, the quantization error is not reduced significantly, as also observed in \cite{FLBeam}. 
	%By contrast, when the bit overhead is smaller than $1$-bit (i.e., the value of $Q$ is less than $3$), the quantization error becomes large, which leads to a poor classification accuracy. 
	%Therefore, we can intuitively conclude that the value of $Q$ does not need to be set to high values such as $4$ or more.

    In Table~\ref{table:Complexity}, we compare the computational complexity of different QCS-based federated learning frameworks. 
    Table~\ref{table:Complexity} shows that FedQCS-EA has a similar complexity order with QCS-QIHT, while providing additional performance gain as can be seen in Fig.~\ref{fig:MNIST}. 
    It is also shown that FedQCS-AE requires a less complexity than FedQCS-EA; this complexity reduction increases with the number of aggregation groups, $G$. 
    In particular, when $G=1$, the complexity of FedQCS-EA is $K$ times lower than that of FedQCS-AE.
    This complexity reduction, however, is attained at the cost of the classification accuracy (see Fig.~\ref{fig:MNIST}), as also discussed in Sec.~\ref{Sec:BussGAMP}. 
    Although QCS-Dither has the lowest complexity order, it suffers from performance degradation in terms of both classification accuracy and NMSE (see Fig.~\ref{fig:MNIST}), while requiring additional signaling overhead as discussed in {\bf Remark 1}. 

	\begin{figure}
		\centering 
		\subfigure[One-bit overhead per entry (i.e., $Q/R = 1$)]
		{\epsfig{file=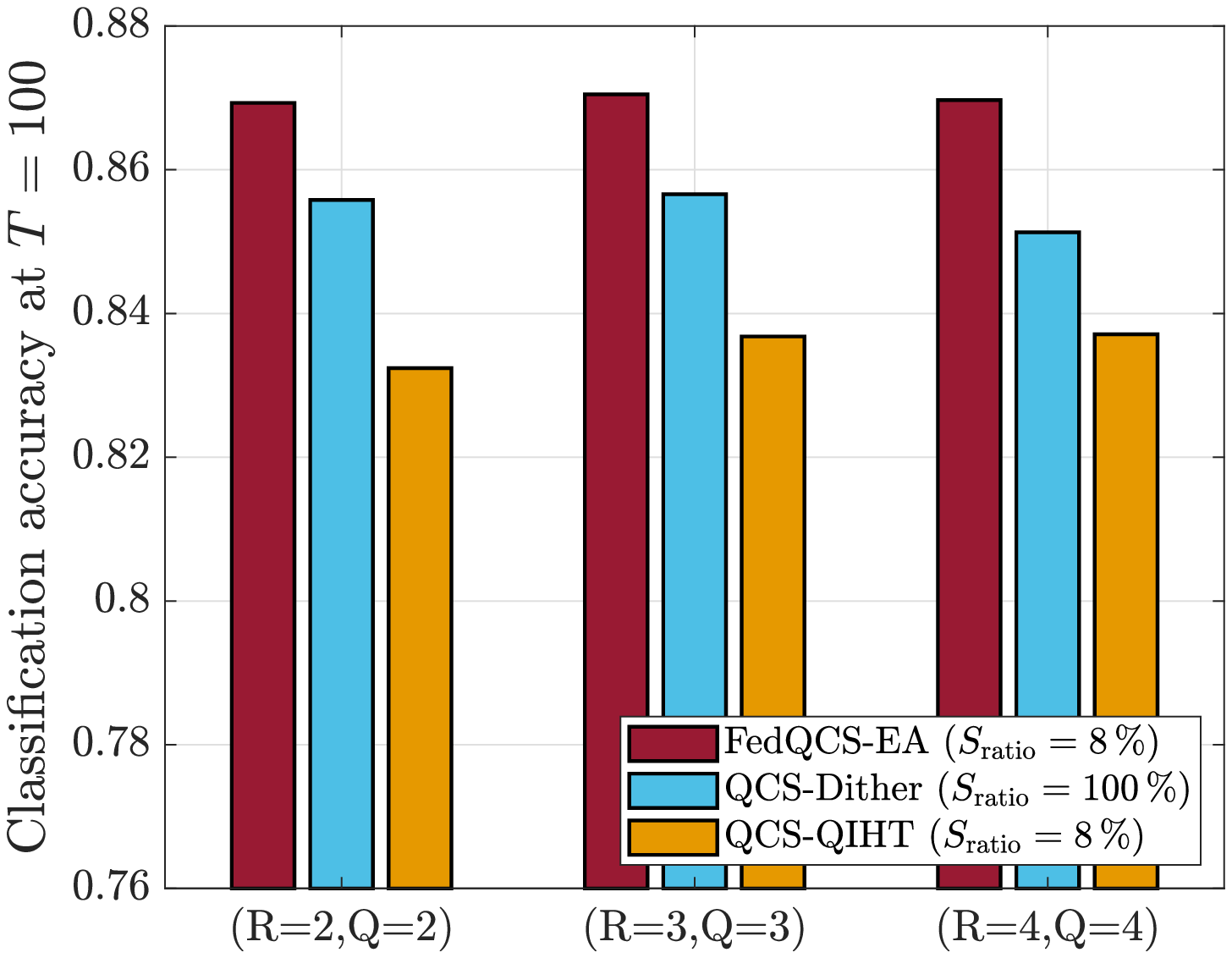, width=7.7cm}}
		%\qquad\qquad\qquad
		\subfigure[$0.5$-bit overhead per entry (i.e., $Q/R = 0.5$)]
		{\epsfig{file=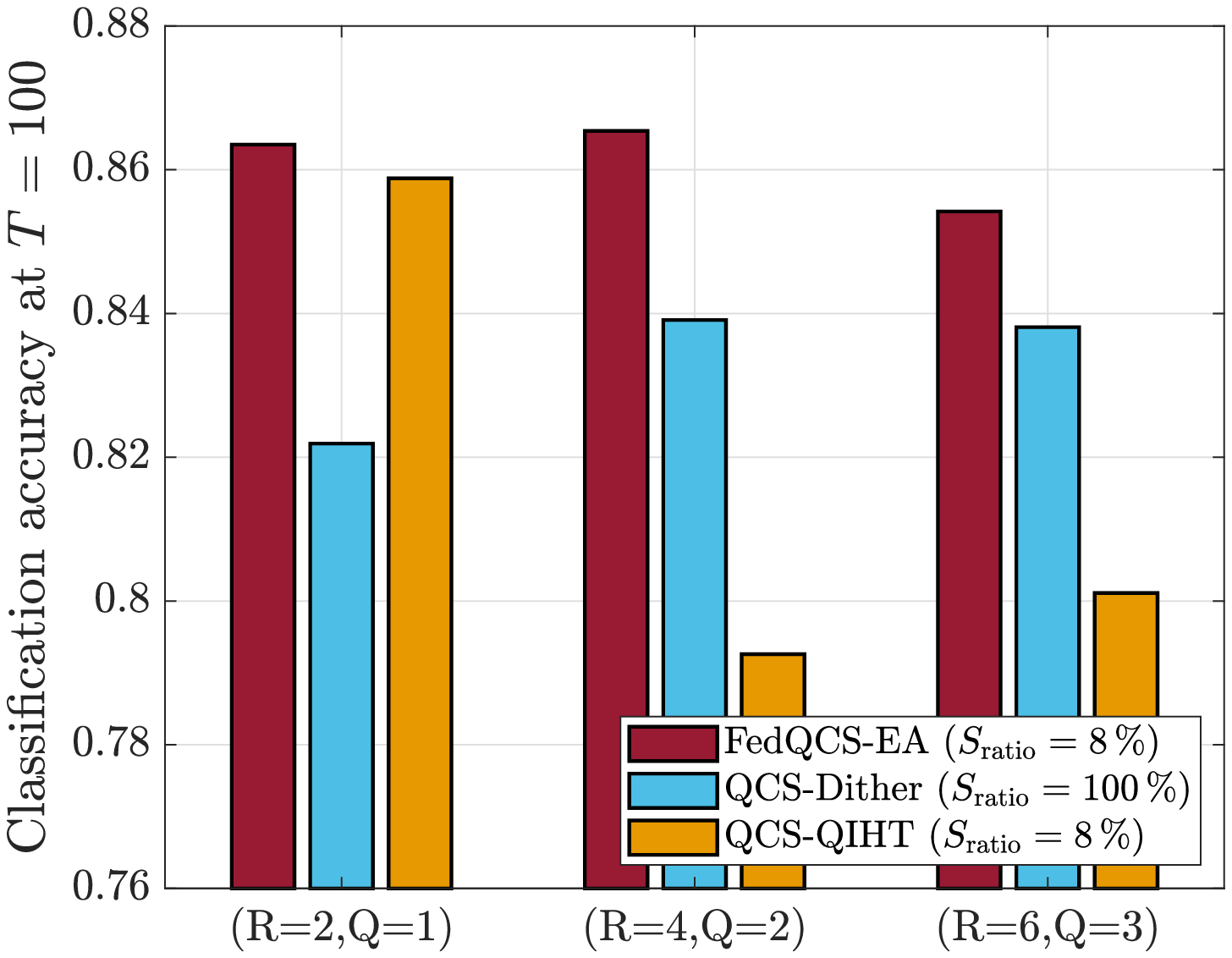, width=7.7cm}} \vspace{-2mm}
		\caption{The classification accuracy of different federated learning frameworks with various choices of $(R,Q)$.} \vspace{-5mm}
		\label{fig:RQ}
	\end{figure}
    
	%\subsection{Effect of Parameters}\label{Sec:Effect of Parameters}
    In Fig.~\ref{fig:RQ}, we evaluate the classification accuracy of different QCS-based federated learning frameworks with various choices of $(R,Q)$.
    %We set $S_{\rm ratio}$ of Q-EM-GAMP and QIHT as $8\,\%$, and that of DithNorm as $100\,\%$.
    Fig.~\ref{fig:RQ} shows that FedQCS-EA achieves the highest classification accuracy regardless of the choice of $(R,Q)$.
    Unlike FedQCS-EA, the performance of the existing QCS-based frameworks changes significantly depending on the choice of $(R,Q)$.
    This result demonstrates the robustness of FedQCS-EA against the changes in $(R,Q)$ compared to the existing QCS-based frameworks. 
    Another interesting observation is that there is the optimal choice of $(R,Q)$ that provides the highest classification accuracy even for the same communication overhead (i.e., even for a fixed ratio $Q/R$).
    %For example, $\{R=3, Q=3\}$ is the optimal pair for the proposed Q-EM-GAMP and DithNorm in our experiments. 
    The reason behind this phenomenon is that the performance of the QCS-based compression depends on two types of errors: {\em compression error} and {\em quantization error}, determined by $R$ and $Q$, respectively. 
    %In particular, the optimal pair is determined by a relative proportion of two error: (i) compression error of compression process and (ii) the error of quantization process. 
    For example, for FedQCS-EA with $Q/R=1$, the performance with $(R,Q) = (2,2)$ is shown to be worse than that with $(R,Q) = (3,3)$ because the effect of the quantization error becomes dominant when $Q=2$. 
    Similarly, the performance with $(R,Q) = (4,4)$ is shown to be worse than that with $(R,Q) = (3,3)$ because the effect of the compression error becomes dominant when $R=4$.
    Therefore, for a fixed communication overhead, a judicious optimization of $R$ and $Q$ is still necessary to maximize the performance of the QCS-based gradient compression. 

	\begin{figure*}
    	\centering
    	{\epsfig{file=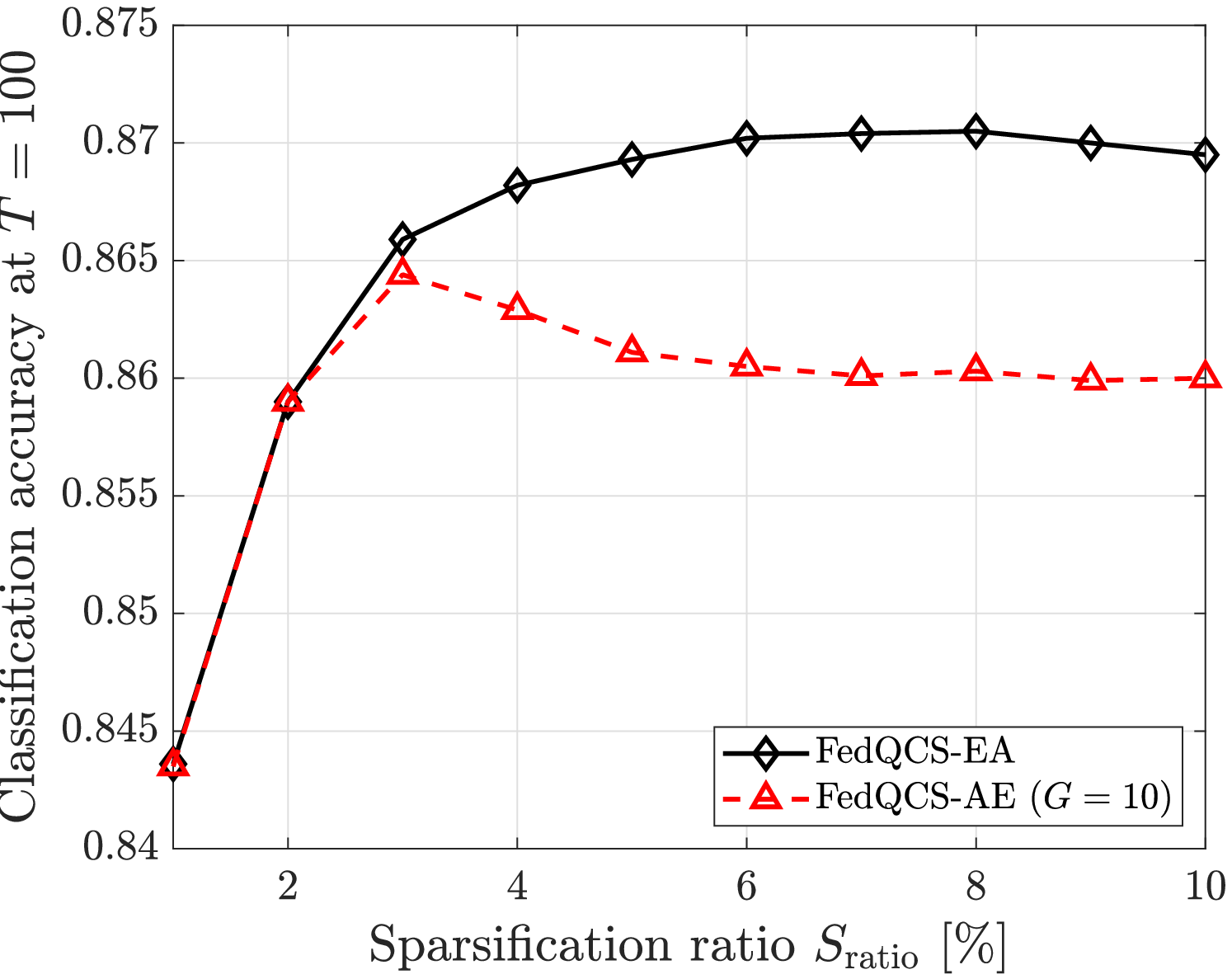, width=7.7cm}} \vspace{-5mm}
    	\caption{The classification accuracy of FedQCS with different values of $S_{\rm ratio}$ when $(R,Q)=(3,3)$.}  \vspace{-5mm}
    	\label{fig:Spars}
    \end{figure*}
    
	%Next, we consider the effectiveness of $S_{\rm ratio}$ when $R=3$, and $Q=3$. 
	In Fig.~\ref{fig:Spars}, we evaluate the classification accuracy of FedQCS with different values of $S_{\rm ratio}$  when $(R,Q)=(3,3)$. 
	Fig.~\ref{fig:Spars} shows that both FedQCS-EA and FedQCS-AE provide the highest classification accuracy at a certain sparsification ratio. 
	The reason behind this phenomenon is that increasing $S_{\rm ratio}$ allows the wireless devices to convey a larger number of gradient entries to the PS, while increasing the number of non-zero entries which leads to degradation in the reconstruction performance at the PS.
	Therefore, there is the trade-off between the amount of the information sent by the devices and the accuracy of the reconstruction at the PS. 
	It is also shown that as $S_{\rm ratio}$ increases, the accuracy of FedQCS-AE degrades earlier than that of the parallel recovery and therefore has a lower value of the optimal sparsification ratio. 
	This coincides with our intuition because the aggregate-and-estimate strategy has a large number of non-zero gradient entries due to the aggregation of the gradient sub-vectors before the estimation, as discussed in Sec.~\ref{Sec:BussGAMP}.
	
% 	Furthermore, the Buss-EM-GAMP adopts the $G=10$, which leads to an increase in the number of non-zero values of gradient compared to the Q-EM-GAMP algorithm which adopts the parallel recovery as mentioned earlier. 
% 	The Q-EM-GAMP algorithm achieves the almost same classification accuracy, about $87\,\%$,  when $S_{\rm ratio}$ is $6\,\%$ to $9\,\%$. These results imply the robustness of Q-EM-GAMP algorithm for $S_{\rm ratio}$. The Buss-EM-GAMP achieves the highest classification accuracy at $S_{\rm ratio} = 3\,\%$. The major reason is that as the $S_{\rm ratio}$ is increases, the number of non-zero values in gradient is also increases. Furthermore, the Buss-EM-GAMP adopts the $G=10$, which leads to an increase in the number of non-zero values of gradient compared to the Q-EM-GAMP algorithm which adopts the parallel recovery as mentioned earlier. 
	%The above result demonstrates that the performance of FedQCS can be maximized by selecting the appropriate value of $S_{\rm ratio}$.

	\section{Conclusion}
	In this paper, we have presented FedQCS, a communication-efficient federated learning framework based on QCS.  
	One prominent feature of FedQCS is that it provides flexible communication overhead that can be made even less than one bit per gradient entry. 
	Another key feature is that gradient reconstruction strategies of FedQCS enable accurate gradient reconstruction at the PS by computing an approximate MMSE estimate of local gradients. 
	By analyzing the reconstruction error as well as the convergence rate of FedQCS, we have demonstrated that the convergence of FedQCS is guaranteed with the rate of $\mathcal{O}\big(\frac{1}{\sqrt{T}}\big)$.
	Using the MNIST dataset, we have demonstrated that FedQCS with one bit overhead per gradient entry suffices to attain almost identical performance as perfect reconstruction with no compression.
	An important direction of future research is to extend the presented framework for the use in wireless multiple access channels which enable further reduction in the communication overhead by allowing the simultaneous transmission of the quantized gradient vectors. 
	In this direction, it would also be important to optimize device scheduling and power control by taking into account different locations and channel conditions of the devices in wireless networks. 
	%It would also be important to study downlink communication of federated learning to enable accurate and efficient broadcasting of the global model to the wireless devices. 

	\appendices
	\section{Proof of Theorem~1}\label{Apdx:Thm1}	
	In this analysis, we omit the index $t$ for the sake of brevity. 
	Consider the asymptotic regime of $N \rightarrow \infty$ and $N/M \rightarrow R$ for a fixed ratio $R\geq 1$. 
	When employing the joint recovery strategy with $G=1$ and large $N$, a gradient reconstruction problem in \eqref{eq:joint_recovery} is formulated as $\tilde{\bf q}_{\mathcal{K},b} = {\bf A}{\bf g}_{\mathcal{K},b} + \tilde{\bf d}_{\mathcal{K},b}$.
%    \begin{align}\label{eq:global_recovery}
%        \tilde{\bf q}_{\mathcal{K},b} = {\bf A}{\bf g}_{\mathcal{K},b} + \tilde{\bf d}_{\mathcal{K},b}.
%    \end{align}
    % where $\tilde{\bf d}_{\mathcal{K},b} =  \sum_{k=1}^K \frac{\rho_k}{\gamma_Q\alpha_{k,b}}  {\bf d}_{k,b}$, and ${\bf g}_{\mathcal{K},b}$ is the $b$-th global gradient sub-vector defined as 
    % \begin{align}
    %     {\bf g}_{{\rm glo},b}
    %     &= \big[ ({\bf g}_{{\rm glo}})_{\mathcal{N}_b(1)}, \cdots, ({\bf g}_{{\rm glo}})_{\mathcal{N}_b(N)}\big]^{\sf T}. 
    % \end{align}		
    As discussed in Sec.~\ref{Sec:BussGAMP}, we have $\tilde{\bf d}_{\mathcal{K},b} \sim \mathcal{N}\big({\bf 0}_{M}, \xi_{\mathcal{K},b} {\bf I}_{M}\big)$ with $\xi_{\mathcal{K},b} =  \kappa_Q\sum_{k=1}^K {\rho_k^2}/{\alpha_{k,b}^2}$ under Assumptions 1 and 2 with large $N$.
    % \begin{align}\label{eq:distortion_global}
    %     \nu_{\mathcal{K},b} = \bigg(\frac{\psi_Q}{\gamma_{Q}^2} -1 \bigg) \sum_{k\in\mathcal{K}} \bigg(\frac{\rho_k}{\alpha_{k,b}}\bigg)^2,
    % \end{align}
    Let $\mu_{{\bf g}_{k,b}}$ and $\nu_{{\bf g}_{k,b}}$ be the mean and the variance of the entry of ${\bf g}_{k,b}$, respectively. 
    Then in the asymptotic regime, we have $1/\alpha_{k,b}^2 = \|{\bf g}_{k,b}\|^2/M   = {\nu_{{\bf g}_{k,b}}} + \mu_{{\bf g}_{k,b}}^2$ under Assumption 1. 
    Utilizing the above facts, the distortion variance $\xi_{\mathcal{K},b}$ is asymptotically given by
    \begin{align}\label{eq:distortion_approx}
        \xi_{\mathcal{K},b} = \kappa_Q \sum_{k=1}^K \rho_k^2\big(\nu_{{\bf g}_{k,b}} + \mu_{{\bf g}_{k,b}}^2 \big)
        = \kappa_Q \big( \tilde{\nu}_{\mathcal{K},b} + \tilde{\mu}_{{\rm sq},\mathcal{K},b}   \big),
    \end{align}
    where $\tilde{\nu}_{\mathcal{K},b} =  \sum_{k=1}^K \rho_k^2\nu_{{\bf g}_{k,b}}$ and $\tilde{\mu}_{{\rm sq},\mathcal{K},b} = \sum_{k=1}^K \rho_k^2\mu_{{\bf g}_{k,b}}^2$.
    Under Assumption 1, the EM-GAMP algorithm applied to estimate the ${\bf g}_{\mathcal{K},b}$ from $\tilde{\bf q}_{\mathcal{K},b}$ exactly behaves like the GAMP algorithm in \cite{Guo:TIT,Bayati:TIT,Rangan:ISIT}. 
    Let $\hat{\bf g}_{\mathcal{K},b}$ be the estimate of ${\bf g}_{\mathcal{K},b}$ obtained by the EM-GAMP algorithm, and $\hat{\bf g}_{\mathcal{K},b}^{\rm LMMSE}$ be the {\em linear} MMSE (LMMSE) estimate of ${\bf g}_{\mathcal{K},b}$ for a given observation $\tilde{\bf q}_{\mathcal{K},b}$. 
    The analysis in \cite{Guo:TIT,Bayati:TIT,Rangan:ISIT} shows that in the asymptotic regime, the GAMP algorithm applied to estimate ${\bf g}_{\mathcal{K},,b}$ from an AWGN observation of ${\bf A}{\bf g}_{\mathcal{K},b}$ is characterized by a scalar state evolution; if this state evolution has a unique fixed point, the solution of the GAMP algorithm converges to the MMSE estimate of ${\bf g}_{\mathcal{K},b}$ for the given observation.
    % Therefore, the estimate of the gradient sub-vector, namely $\hat{\bf g}_{{\rm glo},b}$, obtained by the EM-GAMP algorithm
    The above discussions imply that in the asymptotic regime, the MSE of $\hat{\bf g}_{\mathcal{K},b}$ is lower than the MSE of $\hat{\bf g}_{{\rm LMMSE},b}$, i.e., 
    \begin{align}
        \mathbb{E}\big[\| {\bf g}_{\mathcal{K},b} - \hat{\bf g}_{\mathcal{K},b}\|^2\big]
        \leq
        \mathbb{E}\big[\| {\bf g}_{\mathcal{K},b} -  \hat{\bf g}_{\mathcal{K},b}^{\rm LMMSE}\|^2\big].
        % \overset{(a)}{\leq} \underset{{\bf F} \in \mathbb{R}^{N \times M}}{\min}
        % \mathbb{E}\big[\| {\bf g}_{{\rm glo},b} -  {\bf F} \tilde{\bf q}_{\mathcal{K},b}\|^2\big], 
        \label{eq:LMMSE_bound}
    \end{align}
    %where $\hat{\bf g}_{{\rm LMMSE},b}$ is the {\em linear} MMSE (LMMSE) estimate of ${\bf g}_{{\rm glo},b}$, and the inequality (a) holds because the MSE of the LMMSE estimate cannot be lower than that of the MMSE estimate. 
    Under Assumption 1 with large $N$, $\tilde{\bf d}_{\mathcal{K},b}$ is uncorrelated with ${\bf g}_{\mathcal{K},b}$ by the Bussgang theorem.
    In addition, the variance of ${\bf g}_{\mathcal{K},b}$ is computed as
    $\tilde{\nu}_{\mathcal{K},b}$ under Assumption 1.
    Utilizing these facts, the MSE of $\hat{\bf g}_{\mathcal{K},b}^{\rm LMMSE}$ can be readily computed as \cite{Kay:Book}
	\begin{align}
        \mathbb{E}\big[\| {\bf g}_{\mathcal{K},b} -  \hat{\bf g}_{\mathcal{K},b}^{\rm LMMSE}\|^2\big]
        &= \tilde{\nu}_{\mathcal{K},b} {\sf Tr}\left[  {\bf I}_N  - \tilde{\nu}_{\mathcal{K},b} {\bf A}^{\sf T} \left(  \tilde{\nu}_{\mathcal{K},b}{\bf A}  {\bf A}^{\sf T}  +  \xi_{\mathcal{K},b}{\bf I}_{M}\right)^{-1} {\bf A}
        \right] \nonumber \\
        &\overset{(a)}{=}  \tilde{\nu}_{\mathcal{K},b} N \left( 1 - \frac{\tilde{\nu}_{\mathcal{K},b}}{ R \tilde{\nu}_{\mathcal{K},b} + \xi_{\mathcal{K},b} }\right),
        \label{eq:MSE_LMMSE}
	\end{align}
	where the equality (a) holds because ${\bf A}{\bf A}^{\sf T} = R {\bf I}_M$ in the asymptotic regime from $({\bf A})_{m,n}\sim \mathcal{N}(0,1/M)$.
% 	\begin{align}
%         \mathbb{E}\big[\| {\bf g}_{{\rm glo},b} -  \hat{\bf g}_{{\rm LMMSE},b}\|^2\big]
%         &= \sigma_{{\rm glo},b}^2 {\sf Tr}\left[ \left(  \frac{\sigma_{{\rm glo},b}^2}{\nu_{\mathcal{K},b}} {\bf A} {\bf A}^{\sf T} +  {\bf I}_{M}\right)^{-1}
%         \right] \nonumber \\
%         &\overset{(a)}{=}  \sigma_{{\rm glo},b}^2 {\sf Tr}\left[ \left(  
%         \bigg(\frac{\gamma_{Q}^2}{\psi_Q - \gamma_{Q}^2} \bigg)
%         \bigg(\frac{\sigma_{{\rm glo},b}^2}{ \sigma_{{\rm glo},b}^2 + \tilde{\mu}_{{\rm glo},b}^2 } \bigg) {\bf A} {\bf A}^{\sf T} +  {\bf I}_{M}\right)^{-1}
%         \right],
%         \label{eq:MSE_LMMSE}
% 	\end{align}
% 	where the equality (a) follows from \eqref{eq:distortion_approx}, and $\tilde{\mu}_{{\rm glo},b}^2 = \sum_{k\in\mathcal{K}} \rho_k^2 \mu_{{\bf g},k,b}^2$.
	Since ${\bf g}_{\mathcal{K}}$ is obtained by concatenating the sub-vectors $\{{\bf g}_{\mathcal{K},b}\}_{b=1}^B$, we have
    \begin{align}
        \mathbb{E}\big[\| {\bf g}_{\mathcal{K}} - \hat{\bf g}_{\mathcal{K}}\|^2\big]
        %&=\mathbb{E}\left[\left\| \sum_{k=1}^K \rho_k  \big( {\bf g}_k^{(t)} - \hat{\bf g}_{k}^{(t)} \big)\right\|^2\right] 
        = \sum_{b=1}^B \mathbb{E}\big[\|  {\bf g}_{\mathcal{K},b} -  \hat{\bf g}_{\mathcal{K},b} \|^2\big].  
        \label{eq:MSE_global}
    \end{align}
    Plugging \eqref{eq:distortion_approx}--\eqref{eq:MSE_LMMSE} into \eqref{eq:MSE_global} gives the result in \eqref{eq:Thm1}.

	\section{Proof of Theorem~2}\label{Apdx:Thm2}	
    %This proof follows a similar logic to the convergence proof of a stochastic gradient descent algorithm in \cite{SignSGD}. 
    Under Assumption 3, the improvement of the loss function at iteration $t$ satisfies
    \begin{align}
        F({\bf w}_{t+1}) - F({\bf w}_t)  
        &\leq  \nabla F({\bf w}_t)^{\sf T}({\bf w}_{t+1}-{\bf w}_t)
            +  \frac{\beta}{2}\|{\bf w}_{t+1}-{\bf w}_t\|^2 \nonumber \\
        &\overset{(a)}{=} 
        - \eta_t \nabla F({\bf w}_t)^{\sf T}  \big({\bf g}_{\mathcal{K}}^{(t)} + {\bf e}_t \big)
            +  \eta_t^2\frac{\beta}{2}\|{\bf g}_{\mathcal{K}}^{(t)} + {\bf e}_t  \|^2, \label{eq:loss_improve}
    \end{align}
    where the equality (a) follows from \eqref{eq:global_update} along with ${\bf e}_t =  \hat{\bf g}_{\mathcal{K}}^{(t)} - {\bf g}_{\mathcal{K}}^{(t)}$.
    From the Cauchy–Schwarz inequality, we have $\nabla F({\bf w}_t)^{\sf T} {\bf e}_t \geq - \| \nabla F({\bf w}_t) \| \cdot \|{\bf e}_t\|$ and $2{\bf e}_t^{\sf T} {\bf g}_{\mathcal{K}}^{(t)}  \leq \|{\bf g}_{\mathcal{K}}^{(t)}\|^2 + \|{\bf e}_t\|^2$.
    Applying these inequalities into \eqref{eq:loss_improve} yields
    \begin{align}
        &F({\bf w}_{t+1}) - F({\bf w}_t)  
        \leq 
        - \eta_t \nabla F({\bf w}_t)^{\sf T} {\bf g}_{\mathcal{K}}^{(t)} 
        + \eta_t \| \nabla F({\bf w}_t) \| \cdot \|{\bf e}_t\|
        +  \eta_t^2 {\beta}  \big\{ \|{\bf g}_{\mathcal{K}}^{(t)}\|^2 + \| {\bf e}_t \|^2 \big\}
        \nonumber \\
        &\overset{(b)}{\leq}    
        - \eta_t \nabla F({\bf w}_t)^{\sf T} {\bf g}_{\mathcal{K}}^{(t)} 
        + \eta_t  \sqrt{\epsilon} \| \nabla F({\bf w}_t) \|^2
        +  \eta_t^2 {\beta} \big\{ \|{\bf g}_{\mathcal{K}}^{(t)}\|^2 + \epsilon \| \nabla F({\bf w}_t) \|^2  \big\},
        \label{eq:loss_improve2}
    \end{align}    
    where the inequality (b) holds under Assumption 5.
    Under Assumption 4, we have $\mathbb{E}\big[{\bf g}_{\mathcal{K}}^{(t)} \big| {\bf w}_t \big] = \nabla F ({\bf w}_t)$ and $\mathbb{E}\big[ \| {\bf g}_{\mathcal{K}}^{(t)}\|^2 \big| {\bf w}_t \big] \leq  \|\nabla F ({\bf w}_t) \|^2  + \sigma^2$. 
    Therefore, taking the expectation of both sides of the inequality in \eqref{eq:loss_improve2} conditioned on ${\bf w}_t$ yields
    \begin{align}
        \mathbb{E}\big[  F({\bf w}_{t+1}) - F({\bf w}_t)   \big| {\bf w}_t\big] 
        &\leq 
        - \eta_t ( 1- \sqrt{\epsilon}) \| \nabla F({\bf w}_t) \|^2 
        +  \eta_t^2 {\beta} \big\{ (1 + \epsilon )\| \nabla F({\bf w}_t) \|^2   + \sigma^2\big\}  \nonumber \\
        &= 
         - \eta_t \big\{ ( 1- \sqrt{\epsilon}) -  \eta_t {\beta}  (1 + \epsilon ) \big\} \|\nabla F({\bf w}_t) \|^2 
        +  \eta_t^2 {\beta} \sigma^2.
        \label{eq:loss_improve3}
    \end{align}   
    Plugging a fixed learning rate $\eta_t = \frac{(1-\sqrt{\epsilon})} {2\beta (1+\epsilon) \sqrt{T}}$ into \eqref{eq:loss_improve3} yields  
    \begin{align}
        \mathbb{E}\big[  F({\bf w}_{t+1}) - F({\bf w}_t)   \big| {\bf w}_t\big] 
        &\leq 
          -   \frac{(1-\sqrt{\epsilon})^2} {2\beta (1+\epsilon) \sqrt{T}}   \Big( 1 - \frac{1}{2\sqrt{T}}  \Big) \|\nabla F({\bf w}_t) \|^2 
        +  \frac{(1-\sqrt{\epsilon})^2} {4\beta (1+\epsilon)^2 T } \sigma^2 \nonumber \\
        &\overset{(c)}{\leq} 
          -  \frac{(1-\sqrt{\epsilon})^2} {4\beta (1+\epsilon) \sqrt{T}}  \|\nabla F({\bf w}_t) \|^2 
        +  \frac{(1-\sqrt{\epsilon})^2} {4\beta (1+\epsilon)^2 T } \sigma^2.
        \label{eq:loss_improve4}
    \end{align} 
    where the inequality (c) follows from $\frac{1}{2\sqrt{T}} \leq \frac{1}{2}$.
    By considering a telescoping sum over the iterations, a lower bound of the initial loss with ${\bf w}_1$ is expressed as 
    \begin{align}
        F({\bf w}_{1}) - F({\bf w}^\star) 
        &\geq F({\bf w}_{1})  - \mathbb{E}[F({\bf w}_{T+1})]
        = \sum_{t=1}^{T} \mathbb{E}\left[ F({\bf w}_{t}) - F({\bf w}_{t+1})  \right] \nonumber \\
        &\overset{(d)}{\geq}  \frac{(1-\sqrt{\epsilon})^2} {4\beta (1+\epsilon) } \mathbb{E}\left[ \frac{1}{\sqrt{T}}\sum_{t=1}^{T}  \|\nabla F({\bf w}_t) \|^2  \right] - \frac{(1-\sqrt{\epsilon})^2} {4\beta (1+\epsilon)^2 } \frac{1}{T}\sum_{t=1}^{T} \sigma^2,
        \label{eq:loss_bound}
    \end{align}     
    where the expectation is taken over randomness in the trajectory, and the inequality (d) follows from \eqref{eq:loss_improve4}.
    The inequality in \eqref{eq:loss_bound} can be rewritten as in \eqref{eq:Thm2}, which completes the proof.

\end{document}